# Oxygen Reduction Electrocatalysis with Epitaxially Grown Spinel $MnFe_2O_4$ and $Fe_3O_4$

Alexandria R. C. Bredar[†a], Miles D. Blanchet[‡a], Andricus R. Burton[†], Bethany Matthews[§], Steven R. Spurgeon[§], Ryan B. Comes[‡]*, Byron H. Farnum[†]*

[†]Department of Chemistry and Biochemistry, Auburn University, Auburn, AL 36849
[‡]Department of Physics, Auburn University, Auburn, AL 36849
[§]Energy and Environment Directorate, Pacific Northwest National Laboratory, Richland, WA 99352

[a]Equal Contribution
*Corresponding Authors: ryan.comes@auburn.edu, farnum@auburn.edu

**Abstract**

Nanocrystalline $MnFe_2O_4$ has shown promise as a catalyst for the oxygen reduction reaction (ORR) in alkaline solutions, but the material has been lightly studied as highly ordered thin film catalysts. To examine the role of surface termination and Mn and Fe site occupancy, epitaxial $MnFe_2O_4$ and $Fe_3O_4$ spinel oxide films were grown on (001) and (111) oriented $Nb:SrTiO_3$ perovskite substrates using molecular beam epitaxy and studied as electrocatalysts for the oxygen reduction reaction (ORR). HRXRD and XPS show synthesis of pure phase materials while STEM and RHEED analysis demonstrate island-like growth of (111) surface terminated pyramids on both (001) and (111) oriented substrates, consistent with the literature and attributed to lattice mismatch between the spinel films and perovskite substrate. Cyclic voltammograms under an $N_2$ atmosphere revealed distinct redox features for Mn and Fe surface termination based on comparison of $MnFe_2O_4$ and $Fe_3O_4$. Under $O_2$ atmosphere, electrocatalytic reduction of oxygen was observed at both Mn and Fe redox features; however, diffusion limited current was only achieved at potentials consistent with Fe reduction. This result contrasts with that of nanocrystalline $MnFe_2O_4$ reported in the literature where diffusion limited current is achieved with Mn-based catalysis. This difference is attributed to a low density of Mn surface termination, as determined by the integration of current from CVs collected under $N_2$, in addition to low conductivity through the $MnFe_2O_4$ film due to the degree of inversion. Such low densities are attributed to the synthetic method and island-like growth pattern and highlight challenges in studying ORR catalysis with single-crystal spinel materials.

**Introduction**

The use of fuel cells in our changing energy economy relies heavily on Pt as an oxygen reduction catalyst, which is costly for widespread use of this technology.[1] The exploration of new materials that are cheaper and more abundant is therefore important for the use and expansion of fuel cells. The catalysis of the oxygen reduction reaction (ORR) is complicated by its sluggish kinetics which result from the need for $O_2$ to adsorb to the surface of the catalyst, break the O=O (498 kJ mol$^{-1}$) double bond, and then desorb from the surface, all of which may involve several different peroxide or hydroxide intermediates.[2,3] There are two pathways for ORR in alkaline conditions known as the 4-electron (**Equation 1**) and 2x2-electron pathways (**Equation 2**). The 4-electron pathway is thermodynamically more favorable due to the more efficient conversion of $O_2$ and the avoidance of peroxide side products produced by the 2x2-electron pathway, which can be detrimental to other components of fuels cells.[4] The challenges of finding alternative catalysts to Pt lie in materials that not only have similar overpotentials, but also have similar selectivity for the 4-electron pathway.

1) $O_2 + 2H_2O + 4e^- \rightarrow 4OH^-$     $E^o = 1.23$ V vs RHE
2) $O_2 + H_2O + 2e^- \rightarrow HO_2^- + OH^-$     $E^o = 0.70$ V vs RHE
   $HO_2^- + H_2O + 2e^- \rightarrow 3OH^-$     $E^o = 1.78$ V vs RHE

Spinel oxides have emerged as attractive alternative catalysts for ORR due to their versatility and tunability of their reactivity depending on the chemical nature of their metal cations. Spinel oxides are ternary materials with the chemical formula $AB_2O_4$, where the A cation is typically in the 2+ oxidation state and the B cation is typically in the 3+ oxidation state. Depending on the identity of the metal cations, A and B cations may occupy tetrahedral or octahedral sites ranging from normal ($A^{tet}B^{oct}_2O_4$) to inverse ($A^{oct}B^{tet}B^{oct}O_4$) structures. Many elements can be incorporated into the spinel structure giving a large library of materials to explore.[5]

In terms of low cost and abundance, first-row transition metals have been heavily explored as nanomaterials for ORR, and almost all of the first row transition metal spinels show some propensity towards ORR, with some even demonstrating onset potentials that are competitive with Pt.[6] The ability of these materials to perform ORR lies in the oxidation state promiscuity and the occupation of different coordination sites within the spinel structure. Co, Fe, and Mn for example all have at least two different thermodynamically possible oxidation states that can be utilized when catalyzing the ORR reaction, which is important for charge balance and electron transfer when performing catalysis.[7] Studies of ORR catalysis with spinels have also shown that the occupation of octahedral or tetrahedral sites for different cations can change the catalyst's ability to perform ORR.[8,9] Several studies of ferrite-based spinels ($AFe_2O_4$, where A is Co, Fe, Mn, Ni, or Cu) have been performed with Co and Mn ferrite spinels showing the best ORR activity.[6,10–16] However, differences in synthetic method of nanocrystals and evaluation of ORR make the determination of an outright champion difficult.

The exploration of almost all ORR spinel catalysts has typically been done with nanocrystalline materials on a carbon support. The carbon support is important because it enhances the conductivity and stability of the metal oxide catalyst.[5] However, carbon support materials have also been shown to perform ORR without a transition metal catalyst, albeit less efficiently.[3,17,18] The complexities of carbon support/oxide nanocrystal composites make it difficult to understand the contribution of just the spinel metal oxide to ORR and its true catalytic activity. Understanding the intrinsic thermodynamics and kinetics of the ORR reaction on spinel metal oxide surfaces is important to understanding their catalytic mechanism and realizing their full potential as alternative catalysts to Pt.

To study the surface of a catalyst in greater detail, the use of single crystalline materials is beneficial. ORR catalysis using single crystal Pt, Pd, and Ag have all been achieved, which allows for an understanding of catalysis at a specific surface termination.[19–21] These materials are also readily available as substrates that configure well into electrodes for rotating disk electrochemistry (RDE) experiments typically used to study ORR. Perovskite oxides, such as $LaMO_3$ (where M = Fe, Co, Mn, Ni as a few examples), have also been grown epitaxially and studied for ORR; however, the electrode configuration in most of these studies either prevented RDE experiments from being performed[22,23] or was more relevant to solid oxide fuel cell devices.[24,25] In terms of perovskite oxides, only one study by Kan et al. used RDE to study $La_{0.66}Sr_{0.33}MnO_3$ grown on the conductive perovskite substrate $Nb:SrTiO_3$ and found that diffusion-limited current could not be achieved.[26]

A recent study by Yang et al. explored epitaxial thin film ferrite spinels in an RDE configuration for ORR electrocatalysis. $Fe_3O_4$ and Co-doped $Fe_3O_4$ were grown on MgO substrates and studied electrocatalytically in a configuration where contact was made to the front of the films. While their configuration showed valid RDE measurements with Pd thin films, they were not able to reach diffusion limited current with their spinel films and overall achieved low current densities, most likely due to limitations of lateral charge transport through the spinel thin film. To our knowledge, this is the only other report in the literature which describes ORR electrocatalysis at an epitaxial spinel oxide film.[27]

Herein we describe our study of epitaxially grown $MnFe_2O_4$ and $Fe_3O_4$ for ORR electrocatalysis, which have been studied significantly as nanocrystalline catalysts.[6,11,14,28,29] The epitaxial films were prepared using molecular beam epitaxy (MBE). $MnFe_2O_4$ has been grown epitaxially using MBE previously[30], but not specifically for the study of ORR catalysis. $Nb:SrTiO_3$ (Nb:STO) perovskite substrates were used for oxide growth because of their high conductivity and the current unavailability of conductive spinel substrates. Once grown, the films were made into electrodes for RDE measurements, shown in **Figure 1**.

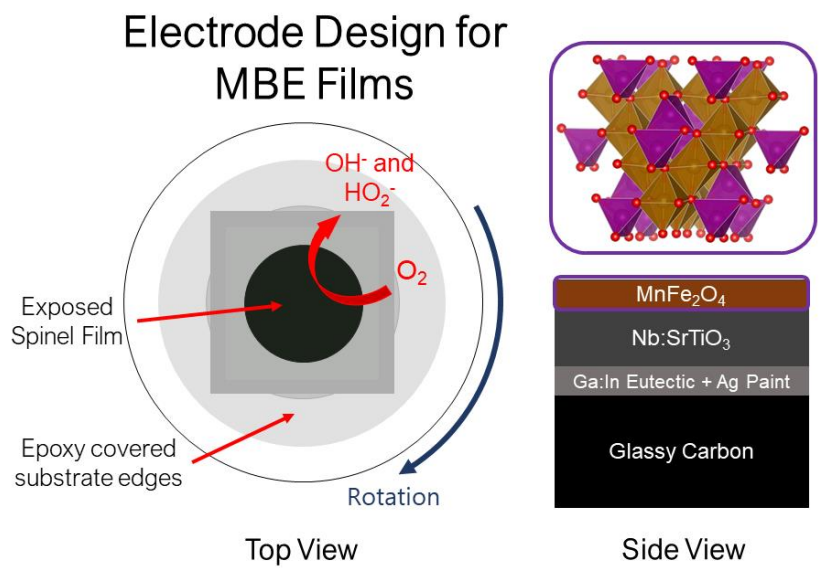

**Figure 1.** Diagram of electrode used to study ferrite spinel epitaxial films for ORR catalysis.

This is the first study to our knowledge of $MnFe_2O_4$ and $Fe_3O_4$ grown on a perovskite substrate used for electrocatalysis and provides insight into the specific mechanism through which $MnFe_2O_4$ and $Fe_3O_4$ catalyze ORR. In this study $MnFe_2O_4$ and $Fe_3O_4$ were grown on (001) and (111) oriented Nb:STO to study the effect of surface termination on film growth and catalysis. Films of varying thickness were also studied. We find that diffusion-limited current consistent with the 4-electron mechanism can be reached in an $O_2$-saturated electrolyte. Interestingly, the overpotential at which this current is achieved (-0.1 V vs. RHE, $\eta_{ORR}$ = 1.33 V) is consistent with the reduction of Fe sites in the crystal structure while catalytic current at Mn sites, occurring at smaller overpotentials (0.65 V, $\eta_{ORR}$ = 0.58 V), was much lower. This result runs counter to those observed for nanocrystalline electrodes where diffusion-limited current is achieved through Mn site catalysis. We hypothesize that this discrepancy is due to a combination of low Mn-surface density on the {111} island facets that form on the perovskite substrate and low conductivity of $MnFe_2O_4$.

**Experimental**

*Synthesis of $MnFe_2O_4$ and $Fe_3O_4$ Films* $MnFe_2O_4$ films were grown on either (001)-oriented or (111)-oriented niobium-doped $SrTiO_3$, (Nb:STO, 0.7 wt%, MTI Crystal) conductive perovskite substrates using molecular beam epitaxy (MBE, Mantis Deposition). Substrates were sonicated in acetone (ACS Grade, VWR) and isopropyl alcohol (ACS Grade, VWR) for ~5 minutes each before being loaded into the MBE chamber. Elemental Mn (99.95%, ACI Alloys) and Fe (99.98%, Sigma-Aldrich) was deposited concurrently during growth while effusion cells were kept at constant temperature, with deposition rates calibrated using a quartz-crystal microbalance (QCM) pre-growth. The sample stage was heated to a constant temperature using an infrared ceramic heating source and measured via a thermocouple on the stage, which causes an overestimation of ~50-100 °C relative to the substrate surface temperature. Samples were grown at 525 °C setpoints and subsequently cooled to ambient temperatures over ~30 minutes. $O_2$ gas was introduced into the chamber and maintained at ~$7.0 \times 10^{-6}$ Torr during growth and cooling of films. $Fe_3O_4$ was grown at the same conditions on (001)-oriented Nb:STO, except $O_2$ gas pressure was maintained at ~$4.5 \times 10^{-6}$ Torr.[31]

*Characterization of $MnFe_2O_4$ and $Fe_3O_4$ Films* Reflection high-energy electron diffraction (RHEED), a technique sensitive to the first few atomic layers of a film surface, was used to monitor the growth process. After growth, samples were analyzed using x-ray photoelectron spectroscopy (XPS, PHI 5400 refurbished by RBD Instruments). The XPS system was connected to the MBE chamber by vacuum transfer line to prevent atmospheric contamination. A monochromatic Al Kα x-ray source, an electron pass energy of 35.75 eV and a scanning step size of 0.05 eV was used for all samples. The conductivity of the Nb:STO substrates negated the need for an electron emission neutralizer for sample charge compensation. All spectra were shifted accordingly to place their Fe $2p_{3/2}$ peaks at 711 eV binding energy.[32,33] Atomic force microscopy (AFM) was used to acquire images of film topography and were acquired using a Park XE7 AFM in non-contact mode. Out-of-plane high resolution X-ray diffraction (HRXRD) and X-ray reflectivity (XRR) was performed using a Rigaku SmartLab system (Cu Kα source) with a hybrid pixel area detector in 0D mode.

Cross-sectional scanning transmission electron microscopy (STEM) samples were prepared using a FEI Helios NanoLab DualBeam $Ga^+$ Focused Ion Beam (FIB) microscope with a standard lift out procedure. STEM high-angle annular dark field (STEM-HAADF) images were collected on a probe-corrected JEOL GrandARM-300F microscope operating at 300 kV, with a convergence semi-angle of 29.7 mrad and a collection angle range of 75–515 mrad. STEM energy-dispersive X-ray spectroscopy (STEM-EDS) maps were acquired using dual JEOL Centurio silicon drift detector setup and the maps shown in Figure X were processed for the Fe *K*, Mn *K*, Sr *L*, and Pt *M* peaks.

***Electrocatalytic studies of MnFe₂O₄ and Fe₃O₄ Films*** Electrodes were constructed from $MnFe_2O_4$ and $Fe_3O_4$ films by mounting films on RDE tips with a glassy carbon (GC) disk working electrode (Pine Research, 5 mm diameter) as the contact material, shown in **Figure 1**. The electrode was constructed using Ga:In eutectic (99.99%, Sigma Aldrich) and silver paint (Ted Pella). A drop of Ga:In eutectic was placed in the middle of the GC electrode to make the electrical contact to the backside of the Nb:STO. Before the film was placed, a ring of Ag paint was placed around the eutectic and was used as a conductive adhesive, so the film would adhere to the glassy carbon surface while keeping the eutectic in place. The film was placed on the eutectic and paint and allowed to dry for 30 minutes. Once the film could not be moved with gentle pressure an inert epoxy (Loctite D-609) was used to cover the film, making sure any exposed GC, eutectic, and Ag paint are covered so they did not interfere with electrochemical measurements. The epoxy was placed such that the edge of the Nb:STO substrate was covered as well, leaving only exposed $MnFe_2O_4$ film. The epoxy was then left to dry at room temperature for at least 24 hours before the electrode was used for electrocatalysis. The final area of the exposed film was measured using ImageJ software.

All electrochemical measurements were performed in 0.1M KOH as the electrolyte. A Hg/HgO reference electrode was used and checked against ferricyanide for every experiment for accurate conversion to RHE, which all potentials are reported against. All cyclic voltammetry (CV) and RDE experiments were performed using a WaveDriver 20 bipotentiostat (Pine Research) with an MSR rotator (Pine Research). For non-catalysis electrochemistry, $N_2$ (UHP 99.999%, Airgas) was purged into the solution for at least 30 minutes. For ORR measurements, $O_2$ (UHP 99.999%, Airgas) was purged into the solution for at least 30 minutes. To confirm our electrode design was applicable for rotation studies, a Pt disk electrode with a similar epoxy coating to our MBE electrodes was tested under ORR conditions.

**Results**

*MBE Synthesis and Characterization* $MnFe_2O_4$ and $Fe_3O_4$ spinel films were grown on 10 mm x 10 mm Nb:STO substrates, then subsequently diced into 5 mm x 5 mm pieces so that different methods of characterization could be performed. Experiments were performed on four different spinel films: 6 nm $MnFe_2O_4$ grown on (001) Nb:STO, 16 nm $MnFe_2O_4$ on (001), 6 nm $MnFe_2O_4$ on (111) Nb:STO, and 21 nm $Fe_3O_4$ on (001) Nb:STO. Initial assurance of successful growth was observed with *in situ* RHEED analysis, which can determine the crystallinity and surface morphology of the material based on the scattering of electrons from the film surface. **Figure 2a-b** shows RHEED analysis of 6 nm (001) and 6 nm (111) $MnFe_2O_4$ for a direct comparison of the impact of substrate orientation. Films grown on (001)-oriented Nb:STO revealed a (001)-oriented spinel structure with an island decorated surface, while those grown on (111)-oriented Nb:STO showed a (111)-oriented spinel structure with smaller island features and a more planar surface quality. These results are indicated by the spotted pattern for $MnFe_2O_4$ grown on the (001) substrate, as opposed to the pattern with vertical streaks combined with spots for the film grown on the (111) substrate. The RHEED pattern for films grown on (001) substrates was the same regardless of thickness, with the RHEED for 16 nm $MnFe_2O_4$ and 21 nm $Fe_3O_4$ on (001) Nb:STO shown in **Figure S1**. For all films used in this study, the RHEED patterns obtained during growth showed little change over time, indicating that the orientations were uniform throughout the epitaxial films. The island growth results are typical for spinel ferrites grown on Nb:STO, where there is a significant lattice mismatch (~7%) between the $MnFe_2O_4$ film lattice parameter (a/2 ≈ 4.24 Å) and that of the substrate (a = 3.905 Å).[31] The {111}-type surfaces are also the minimum energy surface for spinel films[34], which leads to faceting into pyramid-type islands when grown on a (001) substrate. This Volmer-Weber growth mode has been observed previously for $CoFe_2O_4$ films grown on STO.[35]

AFM analysis shown in **Figure 2c-d** further verified the RHEED results, as the root mean square roughness for the 6 nm (001) $MnFe_2O_4$ film was greater at $3.2 \pm 0.5$ nm compared to $1.0 \pm 0.5$ nm for the 6 nm (111) film. AFM for 16 nm (001) $MnFe_2O_4$ and for 21 nm (001) $Fe_3O_4$ is shown in **Figure S1**. The reduction in surface roughness on (111) Nb:STO reflects the smoother surface seen in RHEED, and is also expected given that surface faceting would be reduced since the (111) surface is already the minimum energy surface for spinel oxides. In all cases, residual film strain is fully relaxed to the bulk value due to this island growth mode.

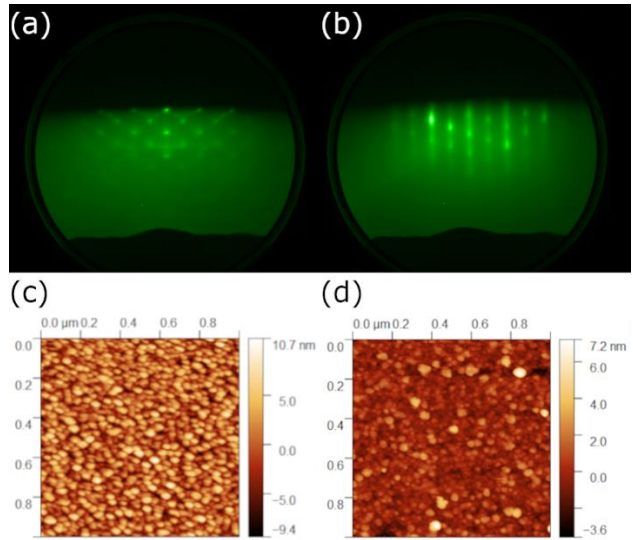

**Figure 2.** RHEED and AFM images of (a,c) 6 nm (001) $MnFe_2O_4$ and (b,d) 6 nm (111) $MnFe_2O_4$.

Out-of-plane HRXRD results for the films are shown in **Figure 3**. $MnFe_2O_4$ ($Fe_3O_4$) grown on (001) substrates showed a peak at 42.5° (43.3°) consistent with a (004) lattice peak of the spinel film, close to the (002) Nb:STO substrate peak at 46.4°. $MnFe_2O_4$ grown on (111) Nb:STO showed a (222) film peak at 36.6°, close to the (111) substrate peak at 40.0°. C-lattice parameters of the $MnFe_2O_4$ ($Fe_3O_4$) film was calculated to be 8.50(1) Å (8.35(1) Å) and the distance between (111) planes of $MnFe_2O_4$ to be 4.91(1) Å. These parameters are consistent with those found in literature for $MnFe_2O_4$ and $Fe_3O_4$ spinels.[36,37] The film thickness for $MnFe_2O_4$ grown on (111) Nb:STO was determined using an XRR fit (**Figure S2**), but this could not be done for films grown on (001) substrates due to their significantly higher surface roughness. Thicknesses of (001)-oriented films were calculated by comparing the composition, QCM calibration rates and growth time between samples, while using the thickness of $MnFe_2O_4$ grown on (111) Nb:STO as a reference. Thicknesses calculated using QCM data have an uncertainty of $\pm 1$ nm.

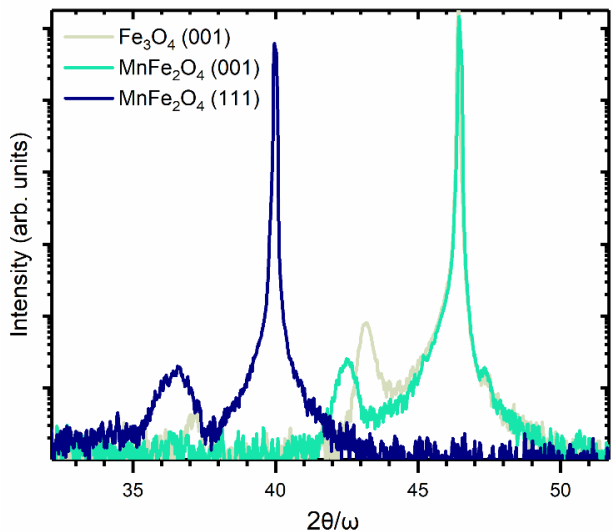

**Figure 3.** HRXRD diffractogram of 21 nm Fe$_3$O$_4$ grown on (001), 16 nm MnFe$_2$O$_4$ grown on (001), and 6 nm grown on (111) oriented Nb:STO substrates. Large peaks at 40.0° and 46.4° indicate the (111) and (002) lattice planes of the Nb:STO substrate. Smaller peaks at 36.6°, 42.5°, and 43.3° are consistent with (222) MnFe$_2$O$_4$, (004) MnFe$_2$O$_4$, and (004) Fe$_3$O$_4$ lattice planes. The vertical axis is on an arbitrary log scale.

STEM analysis of the 6 nm (001) MnFe$_2$O$_4$ film was performed before and after electrochemical experiments to assess possible microstructural and composition changes, as shown in **Figure 4**. Cross-sectional STEM-HAADF images for the film prior to electrochemical experiments revealed a distinct island morphology with clear {111}-type faceting, consistent with RHEED images, and a thin (< 1 nm thick) uniform bridging layer between islands on the substrate surface. STEM-EDS maps of Fe, Mn, Sr, and Pt are shown in **Figure 4c** which reveal a sharp film-substrate interface and uniform island composition. STEM data collected post-catalysis is discussed after electrocatalytic data is presented.

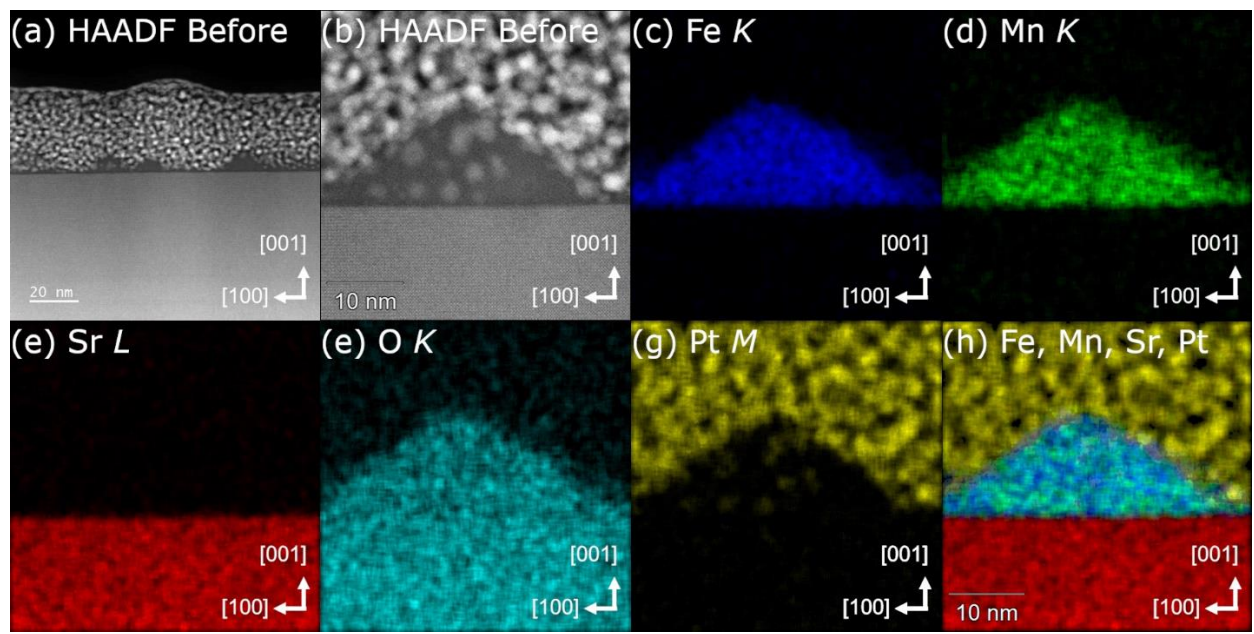

**Figure 4.** Cross-sectional low- and high-magnification STEM-HAADF images of a 6 nm (001)-oriented MnFe$_2$O$_4$ film along with elemental (c-g) and composite (h) STEM-EDS maps collected before cycling.

XPS analysis of the films was performed to characterize their composition, including stoichiometry and metal oxidation states. XPS data for the Mn 2p and Fe 2p regions are shown in **Figure 5** with stoichiometry results presented in **Table 1**. The percentages shown in **Table 1** are based on total metal content excluding oxygen. It should be noted that all spectra were normalized, so the absolute peak intensity is unrelated to stoichiometry. The Mn and Fe stoichiometry for samples still under vacuum was determined by comparing areas of the Mn 2p and Fe 2p regions after implementing a Shirley background subtraction and sensitivity factors of 2.42 and 2.686 for Mn and Fe, respectively. Stoichiometry determination from XPS for samples which were exposed to atmospheric conditions were not considered since the 2p region backgrounds varied greatly due to the effects of atmospheric contamination. The stoichiometry of the films under vacuum varied from sample to sample, with both 6 nm films showing a slight Mn excess, and the 16 nm film showing a slight Mn deficiency. A perfect MnFe$_2$O$_4$ spinel composition should be 33.3% Mn and 66.7% Fe. Values shown in **Table 1** are close to the theoretical estimate; however, reliable stoichiometry determination from XPS is difficult without MnFe$_2$O$_4$ standards. Concentrations determined from XPS using the background shown in **Figure 5** should be considered as an upper-bound for Mn and lower-bound for Fe. EDS analysis was also performed on each film after removal from vacuum to obtain comparative stoichiometry. These data are presented in **Table S1** and show similar estimates to XPS but with slightly smaller Mn at% and larger Fe at%. In terms of oxidation state, all MnFe$_2$O$_4$ samples under vacuum showed similar spectral features, indicating that the oxidation states of the Mn and Fe were consistent across the films. The Mn 2p$_{3/2}$ peaks, located around 640 eV, have satellite peaks around 645 eV, which indicates a Mn$^{2+}$ oxidation state.[38] The Fe 2p$_{1/2}$ peaks of MnFe$_2$O$_4$, positioned around 725 eV, have a satellite peak (~720 eV) which indicates an Fe$^{3+}$ oxidation state.[39] The Fe 2p spectrum of the Fe$_3$O$_4$ film shows a mixed 2+ and 3+ oxidation state, as the Fe$^{3+}$ satellite is less pronounced, consistent with previous reports.[40]

**Table 1: Stoichiometry from XPS**

| MnFe$_2$O$_4$ Film | Mn at% | Fe at% | Mn/Fe |
|---|---|---|---|
| 6 nm (001) Vacuum | 38 | 62 | 0.61 |
| 6 nm (111) Vacuum | 34 | 66 | 0.52 |
| 16 nm (001) Vacuum | 31 | 69 | 0.45 |

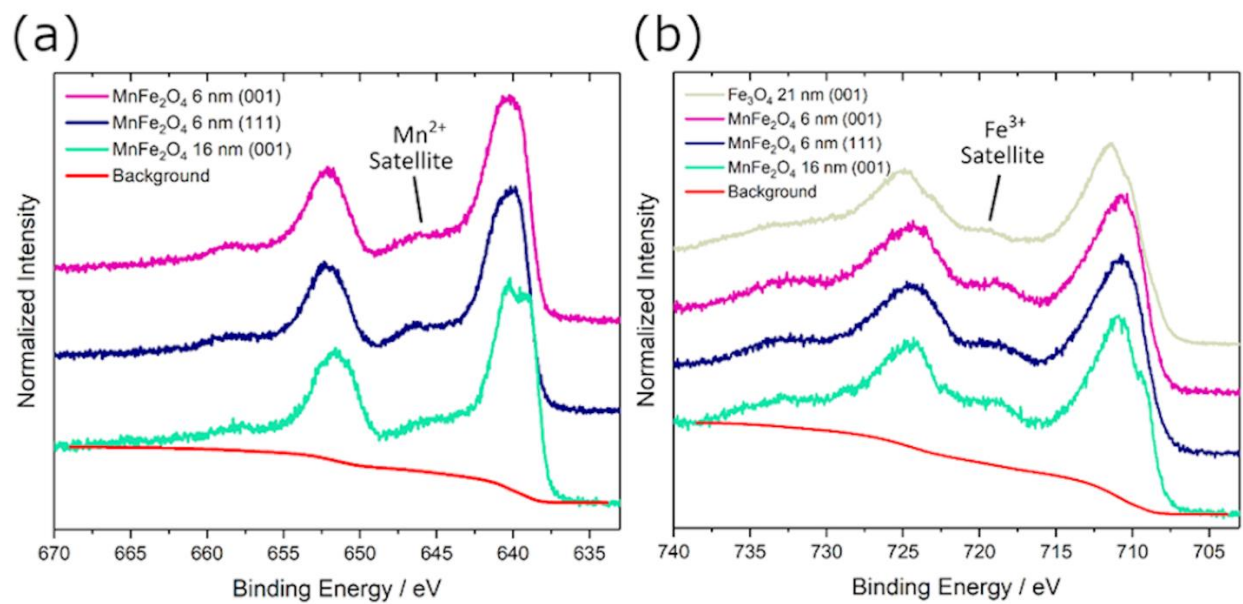

**Figure 5.** XPS of a) Mn 2p region and b) Fe 2p region for 6 nm (001), 6 nm (111) and 16 nm (001) MnFe$_2$O$_4$. Also includes 21 nm (001) Fe$_3$O$_4$.

*ORR Electrocatalysis* Great care was taken in the conversion of $MnFe_2O_4$ and $Fe_3O_4$ films into functional electrodes. Nb:STO substrates with $MnFe_2O_4$ films were mounted on a glassy carbon rotating disk electrode (GC-RDE) using a combination of Ga:In eutectic and Ag paint to make electrical contact between the GC surface and the backside of the Nb:STO substrate. The entire RDE surface was then sealed with epoxy such that only the spinel oxide surface was exposed. Epoxy coverage was found to be important to the stability of the films. Exposure of the Nb:STO edge resulted in rapid degradation of electrochemical features present in the cyclic voltammogram (**Figure S3**) and revealed new XPS peaks in the C 1s region consistent with the presence of Sr and Ti atoms (**Figure S4**). Comparative experiments where the Nb:STO edge was completely covered showed no electrochemical degradation nor the presence of Sr and Ti peaks from XPS. This result indicates that the MBE films are susceptible to etching at the junction of the film and the substrate, thus covering the Nb:STO edge was found to be critical for film stability during electrochemical experiments.

CVs of $MnFe_2O_4$ films were first measured under $N_2$ to understand their basic electrochemical features. These experiments were performed on the exact films which were characterized and discussed in the previous section. **Figure 6a** shows CVs of $MnFe_2O_4$ for the 6 nm (001), 6 nm (111), and 16 nm (001) films. Despite their different surface roughness and thickness the CVs exhibit similar features with two quasi-reversible redox waves at $E_{1/2}$ = 0.96 V and 0.62 V vs RHE. Similar features have been observed in CVs of $MnFe_2O_4$ nanocrystals embedded within carbon black and were assigned to redox chemistry at Mn sites.[11,15] The larger feature at more negative potentials (~0 V vs RHE) has been assigned to redox chemistry at Fe sites.[11,22,41] This is consistent with CV data collected here for the 21 nm (001) $Fe_3O_4$ film which showed no evidence for Mn redox waves and only showed the more negative feature (**Figure 6b**). Scan rate dependent studies under $N_2$ revealed surface-bound redox behavior of both Mn waves (**Figure S5**). The exact assignment of these redox waves is unknown, however literature evidence suggests several possibilities, including assignment of a $Mn^{IV/III}$ wave at 0.92 V and $Mn^{III/II}$ wave at 0.65 V[42], although it has been shown to be difficult to oxidize $Mn^{III} \rightarrow Mn^{IV}$ in the spinel structure.[29] These oxidation states are most likely stabilized by the formation of M-OH or M-O(OH) species at the surface of the $MnFe_2O_4$ film.[43] The second wave (0.65 V) has also been proposed as a $Mn^{II/0}$ wave, which we find unlikely.[11] Studies on bare Nb:STO (**Figure S6**) also indicated there was no contribution from the substrate on the electrochemical features observed in **Figure 6**. Comparing the 6 nm MnFe$_2$O$_4$ films, the (001) film was observed to pass more current than the (111) film, even after the CVs were normalized for the projected surface area. We believe this is consistent with the larger surface roughness observed by AFM for the (001) film compared with the (111) film. Finally, we note that scanning the voltage window to more negative potentials resulted in degradation of the films after repeated cycling. Stable electrochemical results could be obtained if the potential window was thus limited to more positive than -0.5 V. We believe this degradation is due to proton reduction.

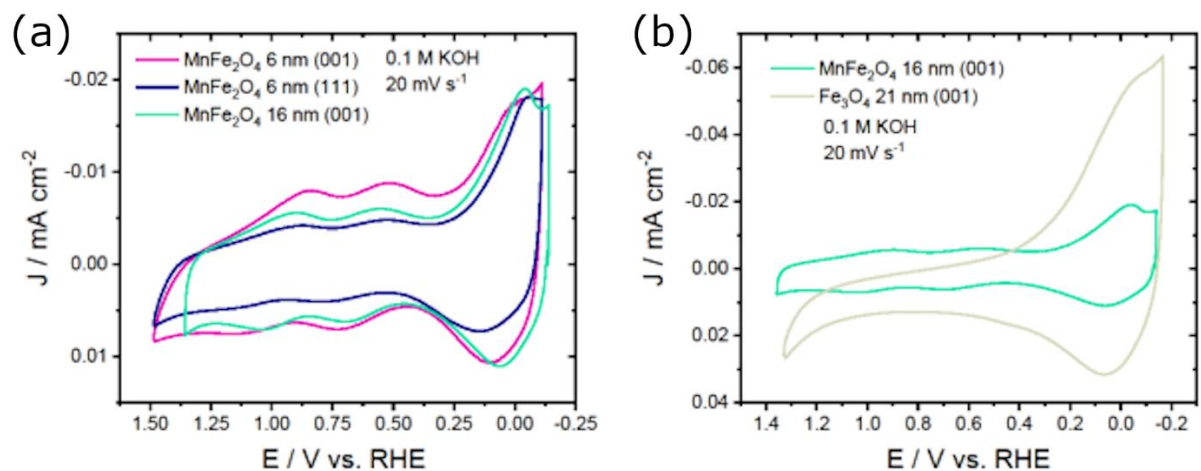

**Figure 6.** a) CVs comparing MnFe$_2$O$_4$ films grown on (001) and (111) substrates. b) CV of 16 nm (001) MnFeO$_2$ under O$_2$ and N$_2$ before RDE experiments.

**Figure 7** shows CVs of the 16 nm (001) MnFe$_2$O$_4$ and 21 nm (001) Fe$_3$O$_4$ film saturated with O$_2$, overlaid with the CV under N$_2$. Under O$_2$ the film shows a significant increase in current with the second Mn feature at 0.62 V, although a true peak for O$_2$ reduction is not observed until 0.03 V vs RHE. Similar data was also observed for 6 nm (001) and (111) films (**Figure S7**). The Fe$_3$O$_4$ film also shows an increase in current upon introduction with O$_2$. There is no early onset in current increase as seen with MnFe$_2$O$_4$, but a similar onset for the O$_2$ reduction peak is observed. The total current passed at the O$_2$ reduction peak for Fe$_3$O$_4$ is notably greater than that observed for the MnFe$_2$O$_4$ samples.

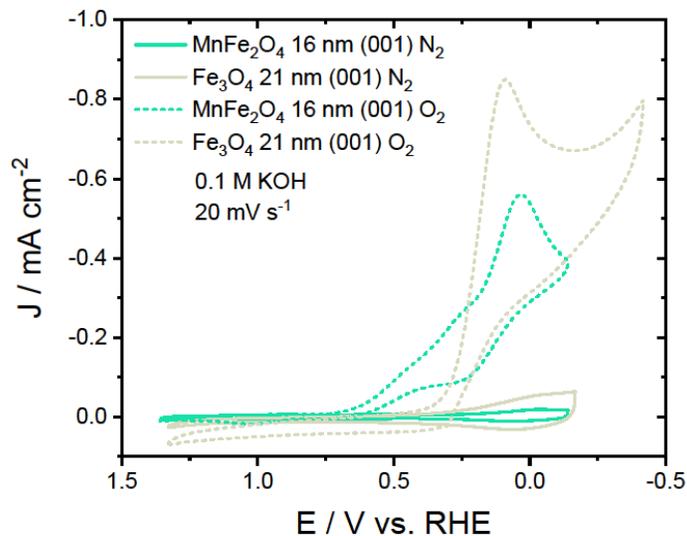

**Figure 7.** Comparison of MnFe$_2$O$_4$ 16 nm (001) and Fe$_3$O$_4$ 21 nm (001) under N$_2$ and O$_2$ atmospheres during cyclic voltammograms.

For a true understanding of the catalytic abilities of MnFe$_2$O$_4$ films, RDE experiments were performed. All films were studied at varying rotation rates to perform a Koutecky-Levich analysis (**Figure 8**). In order to have comparable numbers to other literature reports of MnFe$_2$O$_4$ ORR catalysis, a 20 mV s$^{-1}$ scan rate was used for all rotations. For MnFe$_2$O$_4$ and Fe$_3$O$_4$ at all rotation rates it is apparent that diffusion limited current is being achieved, as evidenced by the plateau in current at potentials < 0 V. From the polarization curves at 1600 rpm (**Figure 8d**), the maximum current observed for 6 nm (001), 6 nm (111), and 16 nm (001) MnFe$_2$O$_4$ films were -5.2 mA cm$^{-2}$, -5.3 mA cm$^{-2}$ and -5.3 mA cm$^{-2}$, respectively, showing that all films pass the same current at the same rotation rate. The E$_{1/2}$ of the main reduction wave, defined as the potential where half the maximum current is achieved for the 1600 rpm condition, was -0.02 V for 6 nm (001), 0 V for 6 nm (111), and 0.04 V for 16 nm (001) MnFe$_2$O$_4$. The onset potential ($E_{onset}$) for any chemical process must be defined at a specific amount of current passed. Typically, the ORR literature has defined this current as 10 µA cm$^{-2}$, which would result in $E_{onset}$ = 0.74, 0.68, and 0.77 V for 6 nm (001), 6 nm (111), and 16 nm (001) films, respectively. Notably, this onset is inconsistent with the large increase in current observed at potentials < 0.25 V. The onset of the large wave was measured for 500 µA cm$^{-2}$ to yield $E_{onset}$ = 0.12 V for 6 nm (001), 0.15 V for 6 nm (111), and 0.22 V for 16 nm (001).

The Fe$_3$O$_4$ film showed similar ORR activity in the Fe region (< 0.25 V) and virtually no catalysis in the Mn region (> 0.25 V). Specifically, at 1600 rpm the maximum current value reached was 5.5 mA cm$^{-2}$ with E$_{onset}$ values of 0.49 V (10 μA cm$^{-2}$) and 0.18 V (500 μA cm$^{-2}$). These results clearly differentiate ORR catalysis based on Fe and Mn metal centers with Fe-based catalysis making a much larger contribution. Complete rotation data for Fe$_3$O$_4$ 21 nm (001) is shown in Figure S8.

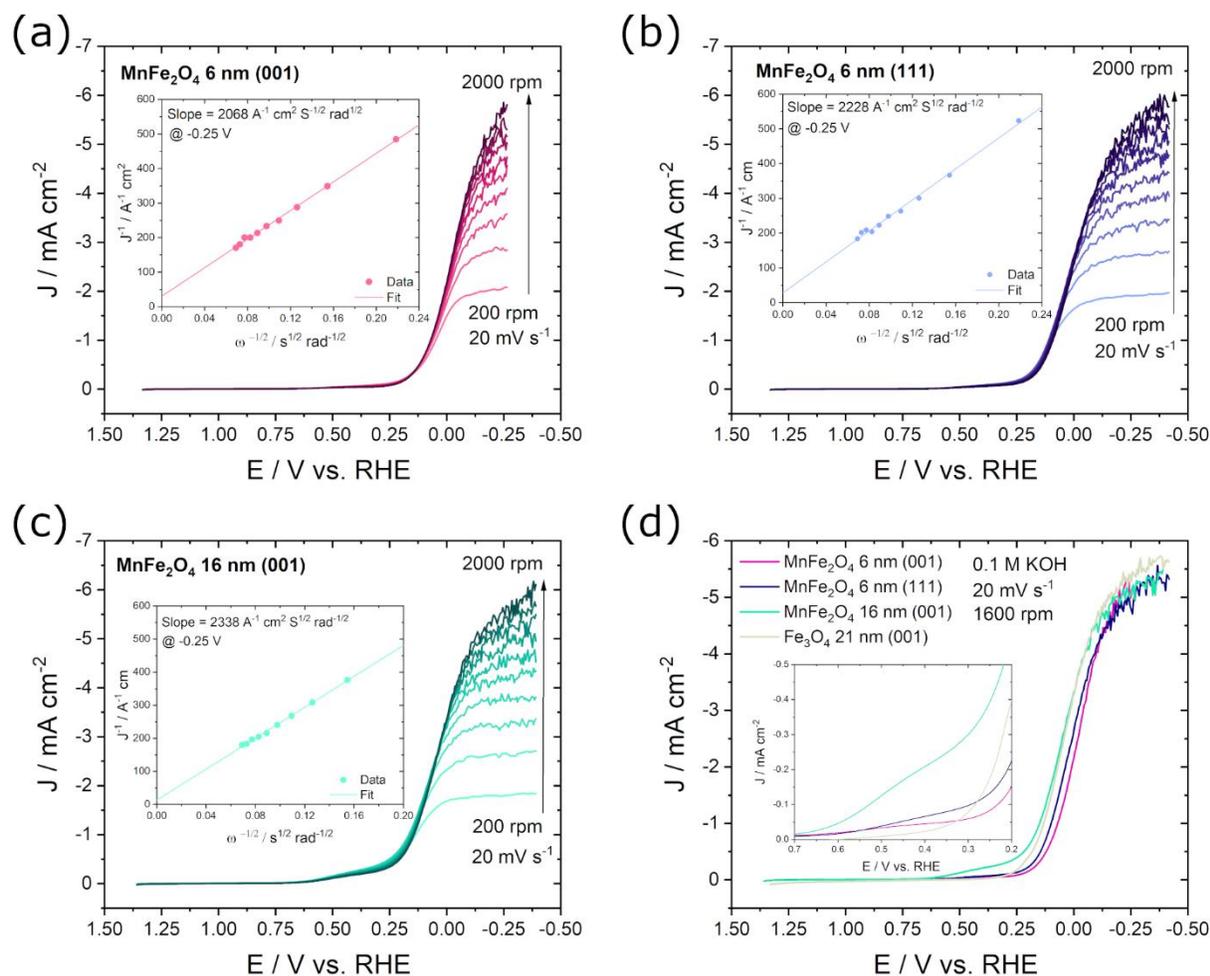

**Figure 8**. Polarization curves at different rotation rates for a) 16 nm (001), b) 6 nm (001), and c) 6 nm (111) MnFe$_2$O$_4$ in 0.1M KOH saturated with O$_2$. K-L plots are inset for each film. d) Overlay of the polarization curve at 1600 RPM for each MnFe$_2$O$_4$ with Fe$_3$O$_4$.

The insets in **Figure 8** show Koutecky-Levich plots generated from RDE data at -0.25 V. At large overpotentials, the kinetic limitations of the system are minimized, and linear trends are observed according to the Koutecky-Levich equation (**Equation 3**). Here, n is the number of electrons transferred at the electrode surface, F is Faraday's constant (96,485 C mol$^{-1}$), $C_0$ is the saturation concentration of $O_2$ in 0.1 M KOH at 1 atm of $O_2$ pressure (1.26 x 10$^{-6}$ mol cm$^{-3}$), $D_0$ is the diffusion coefficient of $O_2$ in 0.1 M KOH (1.93 x 10$^{-5}$ cm$^2$ s$^{-1}$), ν is the kinematic viscosity of the 0.1 M KOH electrolyte (1.09 x 10$^{-2}$ cm$^2$ s$^{-1}$), and ω is the rotation rate in rad s$^{-1}$.[44] Slopes were measured at -0.25 V vs RHE to be 2068, 2228, and 2338 A$^{-1}$ cm$^2$ s$^{1/2}$ rad$^{-1/2}$ for 6 nm (001), 6 nm (111), and 16 nm (001) MnFe$_2$O$_4$, respectively. Using the literature values for $C_0$, $D_0$, and ν, the electron transfer number was calculated from the slope to yield n = 4.2 for 6 nm (001), 3.9 for 6 nm (111), and 3.71 for 16 nm (001). These electron transfer numbers strongly indicate a preference for the direct 4-electron reduction pathway of $O_2$ as opposed to the 2 x 2 electron pathway that produces H$_2$O$_2$ as a byproduct. This is also true for the Fe$_3$O$_4$ film, which had an electron transfer value of 3.91. Complete K-L plots from -0.25 V – 0 V are shown in Figure S9.

$$3)\ \frac{1}{J} = \frac{1}{J_k} + \frac{1}{J_L} = \frac{1}{nFkC_0} + \frac{1}{0.62nFC_0(D_0)^{2/3}\nu^{-1/6}\omega^{1/2}}$$

A Koutecky-Levich analysis was also performed with a Pt disk electrode in the same electrode configuration that was used for the spinel films (i.e. with added epoxy) to validate our electrode construction, as well as have comparison K-L slopes. The data for these studies are shown in **Figure S10**. The calculated electron transfer number calculated for the Pt disk is 4.0.

Following ORR electrocatalysis experiments, the electrolyte was re-purged with N$_2$ to remove O$_2$ from the system and CVs were collected over a wider potential range to investigate the stability of MnFe$_2$O$_4$ films. For all films, two new redox features in the range that O$_2$ reduction appeared concomitant with an increase in non-faradaic current (**Figure S11**). These features are most likely related to oxidation and reduction of Fe sites in the MnFe$_2$O$_4$ structure. Either the exposure to the alkaline conditions of the electrolyte, or the process of performing ORR changed the film surface. In particular, the CV for 16 nm (001) looked significantly different than that collected under N$_2$ prior to electrocatalysis, as a sharp peak arises that is present both under N$_2$ and O$_2$. After completion of these experiments, the 16 nm (001) film was removed from solution for repeat electrochemical analysis. As shown in **Figure S12**, the film was significantly less catalytic in the second electrochemical experiment than it was upon initial electrocatalysis.

XPS, XRD, and STEM characterization of the films was performed post-catalysis to explore how catalysis was affecting the film structure. Due to significant changes in Fe and Mn 2p spectra shapes after exposure to air, stoichiometry of the post-catalysis samples could not reliably be determined. However, from XPS we observed no significant change in the Mn or Fe valence, indicating that the spinel structure of the film remained intact (**Figure S13**). This is further verified by XRD on the films after catalysis which shows no apparent changes in out-of-plane lattice parameters (**Figure S14**). STEM data showed that after catalysis experiments, some changes in the sample morphology are evident, particularly an increase in bridging between islands (**Figure S15**). However, no obvious changes in composition or intermixing occurred.

**Discussion**

The active site for metal oxide ORR catalysis depends significantly on the coordination environment of the metal cation, its oxidation state, and exposure of that active site at the film surface. In several studies of ORR on spinel oxide nanocrystals, investigation of faceting has been performed to understand which crystals faces are best for catalysis.[44–48] For a spinel structure, the surface could consist of two different metals, in two different oxidation states, in two different coordination sites. This can be further complicated if the spinel structure is inverted, so determination of the active cation and coordination is useful for catalyst design. It has been found that the {111} facet of spinels are more active than {001}[46], and we expected this to be reflected in our studies of MBE films by choosing (001) and (111) Nb:STO substrates to dictate surface termination of the spinel. However, due to the lattice mismatch between the spinel films and Nb:STO substrate as well as the lower surface energy of the spinel {111} facets[49], we observe growth of {111} faceted islands on $MnFe_2O_4$ and $Fe_3O_4$ films as evidenced by RHEED and STEM. The island growth pattern observed for $MnFe_2O_4$ and $Fe_3O_4$ spinels is also consistent with what has been observed for $Fe_3O_4$ and $CoFe_2O_4$ growth on a perovskite substrate.[27,35,50] Interestingly, growth of $MnFe_2O_4$ on (111)-oriented Nb:STO substrates also displayed an island growth pattern according to RHEED images but with lower surface roughness than films grown on (001) Nb:STO. Such island growth is likely due to lattice mismatch between the (111) planes of Nb:STO and $MnFe_2O_4$, given that surface energetics no longer drive formation of faceted pyramids along this orientation. The bridges between the islands on the (001) films are (001) faceted, whereas only (111) facets are exposed on the (111) grown $MnFe_2O_4$.

The choice of Nb:STO substrates in this work was important as it allowed for electrical contact to be made with $MnFe_2O_4$ across the entire film area through a backside contact with Nb:STO and the underlying GC electrode. With a non-conductive substrate material such as $MgAl_2O_4$ or MgO, uniform planar growth of $MnFe_2O_4$ (i.e. non-island growth) would be expected; however, electrical contact would need to be made to the front-face of the $MnFe_2O_4$ film. This strategy was recently reported for a series of $Co_xFe_{3-x}O_4$ spinels grown by MBE and studied as ORR catalyst; however, diffusion limited current was not observed.[27] This strategy requires the MBE film to be conductive across the lateral face in order to transport charge from the contact point at the edge of the film to the exposed face in the center. The method of backside contact employed here shows clear diffusion limited current for all electrodes. This suggests that this electrode architecture may be important in the study of MBE spinel catalysts for ORR and emphasizes the need for conductive substrates with better lattice matching to spinel oxides to achieve planar growth of specific surface facets.

From our ORR study of $MnFe_2O_4$ and $Fe_3O_4$ MBE films, a high selectivity for the 4e$^-$ reduction of $O_2$ to OH$^-$ was determined based on electron transfer numbers (n = 3.7 – 4.2) obtained from Koutecky-Levich analyses. The 4e$^-$ reduction process has been reported for $MnFe_2O_4$ nanocrystals by Zhu et al., and Zhou et al.[11,15], however Yang et al. have reported an electron transfer value of only 2.6[6], indicating lower selectivity and more $H_2O_2$ production. Differences in electron transfer numbers notwithstanding, the significant difference between our study and those of nanocrystalline $MnFe_2O_4$ lies is the large overpotential required to achieve diffusion limited current, which when measured at the $E_{1/2}$ for the diffusion limited wave is 0.8 – 1.0 V more negative than what has previously been reported for $MnFe_2O_4$ nanocrystals. These studies have routinely observed $E_{1/2}$ and $E_{onset}$ values at potentials coincident with the Mn-based redox events, as measured under $N_2$. Indeed, we did observe a small amount of ORR catalysis at the Mn waves; however, significant catalytic current was not passed until the Fe-based redox feature was accessible.

We believe the large difference in overpotential between the present study of MBE films and nanocrystalline materials is due to a combination of low density of Mn at the electrode surface and poor conductivity through the film. Further investigation of the CV data collected under $N_2$ provides support for this argument. Integration of both Mn redox waves yields total charge densities of 21.7, 9.0, and 24.3 µC cm$^{-2}$ for the 6 nm (001), 6 nm (111), and 16 nm (001) films, respectively. The assignment of the two Mn redox waves is unclear from the literature of $MnFe_2O_4$; however, we believe they are most likely 1e⁻ processes (e.g. $Mn^{III} \rightarrow Mn^{II}$) corresponding to distinct atomic environments at the electrode surface. Zhou et al. showed that increasing the Mn content from x = 0.5 to 2.5 for $Mn_xFe_{3-x}O_4$ nanocrystals resulted in a decrease in current at the 0.65 V wave and an increase in current at the 0.92 V wave.[15] This behavior suggests two environments of Mn for each redox event as opposed to sequential reduction of the same Mn atom. Therefore, the total charge density can be taken to represent the total Mn density at the surface, assuming 1e⁻ processes. We believe the greater charge densities measured for films grown on (001) vs (111) substrates are a reflection of the larger surface roughness measured by AFM, 3.2 ± 0.5 nm vs 1.0 ± 0.5 nm, respectively.

A calculation of the total surface atomic density (Mn + Fe sites) for the (111) facet of $MnFe_2O_4$ yields a charge density of 154 µC cm⁻² (**Figure S16**, **Table S2**). Comparison of this value with the measured charge densities of $MnFe_2O_4$ films shows the percent coverage of Mn sites to be 14.1, 5.8, and 15.8% for 6 nm (001), 6 nm (111), and 16 nm (001) films, respectively. An ideal (111) surface should be 67% Mn for a normal structure or 33% for an inverse structure, indicating that the MBE films studied here are deficient in available Mn sites needed to catalyze ORR at lower overpotentials.

The question that remains is why are these films deficient in Mn at the surface? XPS data showed Mn:Fe ratios close to 0.5, as would be expected for $MnFe_2O_4$. Therefore, large Mn deficiency throughout the crystal structure should not be the cause. One simple reason could be that the electrode surface is not entirely {111} terminated. STEM results clearly showed {111}-faceted islands separated by regions of (001) bridges. For the STEM sample which was analyzed, islands were as much as 20 nm apart from one another, reducing the exposed $MnFe_2O_4$ surface area by 50-75%. Islands are also known to truncate to (001) surfaces to reduce the overall surface area, which could further reduce the area of {111} facets.[51] This was not directly observed in STEM data for our sample; however, we were only able to observe a limited number of islands.

As alluded to above, the degree of inversion could also be a factor in determining Mn site density as this affects tetrahedral vs octahedral site occupation. In a previous study of $MnFe_2O_4$ grown via MBE, inversion was found to be 20%[30], in agreement with bulk measurements.[52] Although we did not measure the extent of inversion here, based on a value of 20% we would expect the (111) surface to contain 60% Mn distributed over both tetrahedral and octahedral sites.

The measured values reported above are much less than this ideal scenario, suggesting that inversion may result in Mn occupation of octahedral sites which are below the surface. This would yield a {111} surface predominately Fe in character, consistent with the observed electrocatalysis. However, based on the low resolution of the STEM-EDS data along the {111} surface, it is difficult to make a hard conclusion on this claim.

Inversion is also strongly linked to conductivity in spinel ferrites.[52] Yang et al. chose $Fe_3O_4$ to examine spinel MBE ORR electrocatalysis as it is a highly conductive metal oxide due to its inverse structure having a mixture of $Fe^{II}$ and $Fe^{III}$ in octahedral sites which allows for efficient charge hopping through edge-shared octahedra.[31,52] However, introduction of a small amount of $Co^{II}$ into the spinel structure decreased the conductivity of the films significantly, making meaningful elucidation of catalytic properties impossible.[27] The mechanism commonly understood in the literature involves a decrease in the percent inversion such that octahedral sites have a decrease in the mixture of oxidation states (*i.e.* more $Fe^{III}$) which promotes charge hopping.

The CV data collected under $O_2$ without stirring clearly shows that $Fe_3O_4$ passed more current than $MnFe_2O_4$, suggesting that $Fe_3O_4$ is more conductive than $MnFe_2O_4$. Assuming 20% inversion for $MnFe_2O_4$, 90% of octahedral sites would be $Fe^{III}$, resulting in lower conductivity than $Fe_3O_4$ (50% $Fe^{III}$). Interestingly, we note that the overpotential for ORR catalysis and diffusion limited current observed during rotating experiments was very similar between $MnFe_2O_4$ and $Fe_3O_4$, which suggests that communication between surface Fe and the bulk is sufficient to perform catalysis even with Mn in the structure. It is hypothesized that charge hopping in mixed spinels is more likely to occur between cations of the same identity (i.e. $Fe^{III} \rightarrow Fe^{II}$ or $Mn^{III} \rightarrow Mn^{II}$)[31], which supports the consistently high activity of Fe-based catalysis between $Fe_3O_4$ and $MnFe_2O_4$. The low catalytic activity of Mn may therefore be due to lower conductivity to Mn specific sites at the surface which would be controlled by Fe $\rightarrow$ Mn electron transfer. Small differences in inversion and conductivity may also explain the slightly larger catalytic current observed at Mn sites for 16 nm (001) vs 6 nm (001) despite both having a similar Mn site surface density.

For nanocrystalline materials, poor conductivity is overcome by mixing nanocrystals with a conductive carbon material to study ORR reactivity. This has been found to be necessary to achieve lower overpotentials for many different catalysts, including noble metal catalysts and metal oxides.[17,18,53–55] In a study of $Co_3O_4$ spinel by Liang et al., it is very clear that the addition of carbon is necessary for $Co_3O_4$ to be an effective catalyst.[56] There even seems to be a dependence on the type of carbon material used, such as graphene or carbon nanotubes[11], and even the dopant level in the carbon material.[57] Another common component of the nanocrystal/carbon composite is Nafion, an ionic polymer for proton transport, that has an effect on the activity and rate constants associated with catalysis.[58,59] The studies presented here attempt to strip away these factors to focus on the fundamental properties of $MnFe_2O_4$ for ORR catalysis. The use of thin $MnFe_2O_4$ layers deposited on a uniformly conductive substrate was designed to achieve the same result as the nanocrystal/carbon composites, where a highly conductive material is used to transport charge and the oxide surface is responsible for catalysis. Even if the lower surface density of Mn sites is accounted for, the observed catalysis at Mn sites for our MBE films is still lower than what has been observed in $MnFe_2O_4$ nanocrystal/carbon composites, suggesting that carbon supports may play an active role in catalysis.

**Conclusions**

In summary, we have studied $MnFe_2O_4$ and $Fe_3O_4$ films grown via MBE for ORR catalysis. This is the first study of spinel MBE films for ORR catalysis in which diffusion limited current is observed, allowing for determination of the ORR mechanism to be the 4e$^-$ pathway in the case of both materials. We believe this observation was greatly aided by the electrode architecture which employed a conductive Nb:STO substrate for MBE growth, allowing for back-side contact to be made with the spinel films. Importantly, the 4e$^-$ pathway is only observed at potentials consistent with Fe-based redox activity. In the case of $MnFe_2O_4$, catalysis at Mn-based redox features was much lower than what has been observed for nanocrystalline/carbon composite electrodes. We believe this could be the result of multiple factors, including a low density of Mn surface sites and poor conductivity through the $MnFe_2O_4$ film. The low density of Mn surface sites due to island growth highlights the need for conductive, single-crystal spinel substrates for MBE growth. The production of such materials would enable planar spinel catalyst films to be synthesized with control of surface termination (i.e. {111} vs {001}). The studies reported here are among the first of their kind and further research in this area will continue to gain definitive knowledge on the ORR activity of spinel oxides.

## ASSOCIATED CONTENT

**Supporting Information** RHEED, AFM, XRR, EDS and XPS characterization of epitaxial films. Cyclic voltammograms and Koutecky-Levich analysis of ORR electrocatalysis. Surface charge density calculations.

## Acknowledgments

A.R.C.B, M.D.B., A.R.B., R.B.C, and B.H.F. acknowledge support from the National Science Foundation Division of Materials Research through grant NSF-DMR-1809847. Additionally, A.R.C.B, M.D.B., and A.R.B. acknowledge support from the Alabama EPSCOR Graduate Research Scholars program. HRXRD was performed with a Rigaku SmartLab instrument purchased with support from the National Science Foundation Major Research Instrumentation program through grant NSF-DMR-2018794. B.M. and S.R.S. acknowledge support from a Chemical Dynamics Initiative (CDi) Laboratory Directed Research and Development (LDRD) project at Pacific Northwest National Laboratory (PNNL). PNNL is a multiprogram national laboratory operated for the U.S. Department of Energy (DOE) by Battelle Memorial Institute under Contract No. DE-AC05-76RL0-1830. STEM sample preparation was performed in the Environmental Molecular Sciences Laboratory (EMSL), a national scientific user facility sponsored by the Department of Energy's Office of Biological and Environmental Research and located at PNNL. STEM imaging was performed in the Radiological Microscopy Suite (RMS), located in the Radiochemical Processing Laboratory (RPL) at PNNL.

# References

(1) Gasteiger, H. A.; Marković, N. M. Just a Dream—or Future Reality? *Science* **2009**, *324* (5923), 48–49. https://doi.org/10.1126/science.1172083.
(2) Ge, X.; Sumboja, A.; Wuu, D.; An, T.; Li, B.; Goh, F. W. T.; Hor, T. S. A.; Zong, Y.; Liu, Z. Oxygen Reduction in Alkaline Media: From Mechanisms to Recent Advances of Catalysts. *ACS Catal.* **2015**, *5* (8), 4643–4667. https://doi.org/10.1021/acscatal.5b00524.
(3) Ma, R.; Lin, G.; Zhou, Y.; Liu, Q.; Zhang, T.; Shan, G.; Yang, M.; Wang, J. A Review of Oxygen Reduction Mechanisms for Metal-Free Carbon-Based Electrocatalysts. *Npj Comput. Mater.* **2019**, *5* (1), 1–15. https://doi.org/10.1038/s41524-019-0210-3.
(4) Cheng, F.; Chen, J. Metal–Air Batteries: From Oxygen Reduction Electrochemistry to Cathode Catalysts. *Chem. Soc. Rev.* **2012**, *41* (6), 2172–2192. https://doi.org/10.1039/C1CS15228A.
(5) Zhao, Q.; Yan, Z.; Chen, C.; Chen, J. Spinels: Controlled Preparation, Oxygen Reduction/Evolution Reaction Application, and Beyond. *Chem. Rev.* **2017**, *117* (15), 10121–10211. https://doi.org/10.1021/acs.chemrev.7b00051.
(6) Yang, Y.; Xiong, Y.; Holtz, M. E.; Feng, X.; Zeng, R.; Chen, G.; DiSalvo, F. J.; Muller, D. A.; Abruña, H. D. Octahedral Spinel Electrocatalysts for Alkaline Fuel Cells. *Proc. Natl. Acad. Sci.* **2019**, *116* (49), 24425–24432. https://doi.org/10.1073/pnas.1906570116.
(7) Vielstich, W.; Lamm, A.; Gasteiger, H. Handbook of Fuel Cells. Fundamentals, Technology, Applications. **2003**.
(8) Wu, G.; Wang, J.; Ding, W.; Nie, Y.; Li, L.; Qi, X.; Chen, S.; Wei, Z. A Strategy to Promote the Electrocatalytic Activity of Spinels for Oxygen Reduction by Structure Reversal. *Angew. Chem. Int. Ed.* **2016**, *55* (4), 1340–1344. https://doi.org/10.1002/anie.201508809.
(9) Zhou, Y.; Xi, S.; Wang, J.; Sun, S.; Wei, C.; Feng, Z.; Du, Y.; Xu, Z. J. Revealing the Dominant Chemistry for Oxygen Reduction Reaction on Small Oxide Nanoparticles. *ACS Catal.* **2018**, *8* (1), 673–677. https://doi.org/10.1021/acscatal.7b03864.
(10) Wang, C.; Daimon, H.; Sun, S. Dumbbell-like Pt−$Fe_3O_4$ Nanoparticles and Their Enhanced Catalysis for Oxygen Reduction Reaction. *Nano Lett.* **2009**, *9* (4), 1493–1496. https://doi.org/10.1021/nl8034724.
(11) Zhu, H.; Zhang, S.; Huang, Y.-X.; Wu, L.; Sun, S. Monodisperse $M_xFe_{3-x}O_4$ (M = Fe, Cu, Co, Mn) Nanoparticles and Their Electrocatalysis for Oxygen Reduction Reaction. *Nano Lett.* **2013**, *13* (6), 2947–2951. https://doi.org/10.1021/nl401325u.
(12) Li, P.; Ma, R.; Zhou, Y.; Chen, Y.; Zhou, Z.; Liu, G.; Liu, Q.; Peng, G.; Liang, Z.; Wang, J. In Situ Growth of Spinel $CoFe_2O_4$ Nanoparticles on Rod-like Ordered Mesoporous Carbon for Bifunctional Electrocatalysis of Both Oxygen Reduction and Oxygen Evolution. *J. Mater. Chem. A* **2015**, *3* (30), 15598–15606. https://doi.org/10.1039/C5TA02625C.
(13) Zhao, X.; Fu, Y.; Wang, J.; Xu, Y.; Tian, J.-H.; Yang, R. Ni-Doped $CoFe_2O_4$ Hollow Nanospheres as Efficient Bi-Functional Catalysts. *Electrochimica Acta* **2016**, *201*, 172–178. https://doi.org/10.1016/j.electacta.2016.04.001.
(14) Khilari, S.; Pradhan, D. $MnFe_2O_4$@nitrogen-Doped Reduced Graphene Oxide Nanohybrid: An Efficient Bifunctional Electrocatalyst for Anodic Hydrazine Oxidation and Cathodic Oxygen Reduction. *Catal. Sci. Technol.* **2017**, *7* (24), 5920–5931. https://doi.org/10.1039/C7CY01844D.
(15) Zhou, Y.; Xi, S.; Wang, J.; Sun, S.; Wei, C.; Feng, Z.; Du, Y.; Xu, Z. J. Revealing the Dominant Chemistry for Oxygen Reduction Reaction on Small Oxide Nanoparticles. *ACS Catal.* **2018**, *8* (1), 673–677. https://doi.org/10.1021/acscatal.7b03864.
(16) Xiong, Y.; Yang, Y.; Feng, X.; DiSalvo, F. J.; Abruña, H. D. A Strategy for Increasing the Efficiency of the Oxygen Reduction Reaction in Mn-Doped Cobalt Ferrites. *J. Am. Chem. Soc.* **2019**, *141* (10), 4412–4421. https://doi.org/10.1021/jacs.8b13296.
(17) Poux, T.; Napolskiy, F. S.; Dintzer, T.; Kéranguéven, G.; Istomin, S. Ya.; Tsirlina, G. A.; Antipov, E. V.; Savinova, E. R. Dual Role of Carbon in the Catalytic Layers of Perovskite/Carbon

Composites for the Electrocatalytic Oxygen Reduction Reaction. *Catal. Today* **2012**, *189* (1), 83–92. https://doi.org/10.1016/j.cattod.2012.04.046.

(18) Kim, J. H.; Cheon, J. Y.; Shin, T. J.; Park, J. Y.; Joo, S. H. Effect of Surface Oxygen Functionalization of Carbon Support on the Activity and Durability of Pt/C Catalysts for the Oxygen Reduction Reaction. *Carbon* **2016**, *101*, 449–457. https://doi.org/10.1016/j.carbon.2016.02.014.

(19) Markovic, N. M.; Gasteiger, H. A.; Ross, P. N. Oxygen Reduction on Platinum Low-Index Single-Crystal Surfaces in Sulfuric Acid Solution: Rotating Ring-Pt(Hkl) Disk Studies. *J. Phys. Chem.* **1995**, *99* (11), 3411–3415. https://doi.org/10.1021/j100011a001.

(20) Climent, V.; Marković, N. M.; Ross, P. N. Kinetics of Oxygen Reduction on an Epitaxial Film of Palladium on Pt(111),. *J. Phys. Chem. B* **2000**, *104* (14), 3116–3120. https://doi.org/10.1021/jp993480t.

(21) Blizanac, B. B.; Ross, P. N.; Marković, N. M. Oxygen Reduction on Silver Low-Index Single-Crystal Surfaces in Alkaline Solution:  Rotating Ring DiskAg(Hkl) Studies. *J. Phys. Chem. B* **2006**, *110* (10), 4735–4741. https://doi.org/10.1021/jp056050d.

(22) Stoerzinger, K. A.; Lü, W.; Li, C.; Ariando; Venkatesan, T.; Shao-Horn, Y. Highly Active Epitaxial $La_{(1-x)}Sr_xMnO_3$ Surfaces for the Oxygen Reduction Reaction: Role of Charge Transfer. *J. Phys. Chem. Lett.* **2015**, *6* (8), 1435–1440. https://doi.org/10.1021/acs.jpclett.5b00439.

(23) May, K. J.; Fenning, D. P.; Ming, T.; Hong, W. T.; Lee, D.; Stoerzinger, K. A.; Biegalski, M. D.; Kolpak, A. M.; Shao-Horn, Y. Thickness-Dependent Photoelectrochemical Water Splitting on Ultrathin $LaFeO3$ Films Grown on $Nb:SrTiO_3$. *J. Phys. Chem. Lett.* **2015**, *6* (6), 977–985. https://doi.org/10.1021/acs.jpclett.5b00169.

(24) Lee, D.; Grimaud, A.; Crumlin, E. J.; Mezghani, K.; Habib, M. A.; Feng, Z.; Hong, W. T.; Biegalski, M. D.; Christen, H. M.; Shao-Horn, Y. Strain Influence on the Oxygen Electrocatalysis of the (100)-Oriented Epitaxial $La_2NiO_{4+\delta}$ Thin Films at Elevated Temperatures. *J. Phys. Chem. C* **2013**, *117* (37), 18789–18795. https://doi.org/10.1021/jp404121p.

(25) Lee, D.; Jacobs, R.; Jee, Y.; Seo, A.; Sohn, C.; Ievlev, A. V.; Ovchinnikova, O. S.; Huang, K.; Morgan, D.; Lee, H. N. Stretching Epitaxial $La_{0.6}Sr_{0.4}CoO_{3-\delta}$ for Fast Oxygen Reduction. *J. Phys. Chem. C* **2017**, *121* (46), 25651–25658. https://doi.org/10.1021/acs.jpcc.7b06374.

(26) Kan, D.; Orikasa, Y.; Nitta, K.; Tanida, H.; Kurosaki, R.; Nishimura, T.; Sasaki, T.; Guo, H.; Ozaki, Y.; Uchimoto, Y.; Shimakawa, Y. Overpotential-Induced Introduction of Oxygen Vacancy in $La_{0.67}Sr_{0.33}MnO_3$ Surface and Its Impact on Oxygen Reduction Reaction Catalytic Activity in Alkaline Solution. *J. Phys. Chem. C* **2016**, *120* (11), 6006–6010. https://doi.org/10.1021/acs.jpcc.5b11664.

(27) Yang, Y.; Zeng, R.; Paik, H.; Kuo, D.-Y.; Schlom, D. G.; DiSalvo, F. J.; Muller, D. A.; Suntivich, J.; Abruña, H. D. Epitaxial Thin-Film Spinel Oxides as Oxygen Reduction Electrocatalysts in Alkaline Media. *Chem. Mater.* **2021**, *33* (11), 4006–4013. https://doi.org/10.1021/acs.chemmater.1c00388.

(28) Li, Z.; Gao, K.; Han, G.; Wang, R.; Li, H.; Zhao, X. S.; Guo, P. Solvothermal Synthesis of $MnFe_2O_4$ Colloidal Nanocrystal Assemblies and Their Magnetic and Electrocatalytic Properties. *New J. Chem.* **2014**, *39* (1), 361–368. https://doi.org/10.1039/C4NJ01466A.

(29) Wei, C.; Feng, Z.; Baisariyev, M.; Yu, L.; Zeng, L.; Wu, T.; Zhao, H.; Huang, Y.; Bedzyk, M. J.; Sritharan, T.; Xu, Z. J. Valence Change Ability and Geometrical Occupation of Substitution Cations Determine the Pseudocapacitance of Spinel Ferrite $XFe_2O_4$ (X = Mn, Co, Ni, Fe). *Chem. Mater.* **2016**, *28* (12), 4129–4133. https://doi.org/10.1021/acs.chemmater.6b00713.

(30) Matzen, S.; Moussy, J.-B.; Mattana, R.; Bouzehouane, K.; Deranlot, C.; Petroff, F.; Cezar, J. C.; Arrio, M.-A.; Sainctavit, Ph.; Gatel, C.; Warot-Fonrose, B.; Zheng, Y. Epitaxial Growth and Ferrimagnetic Behavior of $MnFe_2O_4$ (111) Ultrathin Layers for Room-Temperature Spin Filtering. *Phys. Rev. B* **2011**, *83* (18), 184402. https://doi.org/10.1103/PhysRevB.83.184402.

(31) Bhargava, A.; Eppstein, R.; Sun, J.; Smeaton, M. A.; Paik, H.; Kourkoutis, L. F.; Schlom, D. G.; Caspary Toroker, M.; Robinson, R. D. Breakdown of the Small-Polaron Hopping Model in

(32) Rajagiri, P.; Sahu, B. N.; Venkataramani, N.; Prasad, S.; Krishnan, R. Effect of Substrate Temperature on Magnetic Properties of MnFe$_2$O$_4$ Thin Films. *AIP Adv.* **2018**, *8* (5), 056112. https://doi.org/10.1063/1.5007792.

(33) Thapa, S.; Paudel, R.; Blanchet, M. D.; Gemperline, P. T.; Comes, R. B. Probing Surfaces and Interfaces in Complex Oxide Films via in Situ X-Ray Photoelectron Spectroscopy. *J. Mater. Res.* **2020**, *36*, 26-51. https://doi.org/10.1557/s43578-020-00070-9.

(34) Zheng, H.; Zhan, Q.; Zavaliche, F.; Sherburne, M.; Straub, F.; Cruz, M. P.; Chen, L.-Q.; Dahmen, U.; Ramesh, R. Controlling Self-Assembled Perovskite−Spinel Nanostructures. *Nano Lett.* **2006**, *6* (7), 1401–1407. https://doi.org/10.1021/nl060401y.

(35) Comes, R.; Gu, M.; Khokhlov, M.; Liu, H.; Lu, J.; Wolf, S. A. Electron Molecular Beam Epitaxy: Layer-by-Layer Growth of Complex Oxides via Pulsed Electron-Beam Deposition. *J. Appl. Phys.* **2013**, *113* (2), 023303-023303–023307. https://doi.org/doi:10.1063/1.4774238.

(36) Permien, S.; Hain, H.; Scheuermann, M.; Mangold, S.; Mereacre, V.; Powell, A. K.; Indris, S.; Schürmann, U.; Kienle, L.; Duppel, V.; Harm, S.; Bensch, W. Electrochemical Insertion of Li into Nanocrystalline MnFe2O4: A Study of the Reaction Mechanism. *RSC Adv.* **2013**, *3* (45), 23001–23014. https://doi.org/10.1039/C3RA44383C.

(37) Zhuang, L.; Zhang, W.; Zhao, Y.; Shen, H.; Lin, H.; Liang, J. Preparation and Characterization of Fe3O4 Particles with Novel Nanosheets Morphology and Magnetochromatic Property by a Modified Solvothermal Method. *Sci. Rep.* **2015**, *5* (1), 9320. https://doi.org/10.1038/srep09320.

(38) Biesinger, M. C.; Payne, B. P.; Grosvenor, A. P.; Lau, L. W. M.; Gerson, A. R.; Smart, R. St. C. Resolving Surface Chemical States in XPS Analysis of First Row Transition Metals, Oxides and Hydroxides: Cr, Mn, Fe, Co and Ni. *Appl. Surf. Sci.* **2011**, *257* (7), 2717–2730. https://doi.org/10.1016/j.apsusc.2010.10.051.

(39) Yamashita, T.; Hayes, P. Analysis of XPS Spectra of Fe$^{2+}$ and Fe$^{3+}$ Ions in Oxide Materials. *Appl. Surf. Sci.* **2008**, *254* (8), 2441–2449. https://doi.org/10.1016/j.apsusc.2007.09.063.

(40) Chambers, S. A.; Joyce, S. A. Surface Termination, Composition and Reconstruction of Fe$_3$O$_4$(001) and γ-Fe$_2$O$_3$(001). *Surf. Sci.* **1999**, *420* (2), 111–122. https://doi.org/10.1016/S0039-6028(98)00657-8.

(41) Sander, M.; Hofstetter, T. B.; Gorski, C. A. Electrochemical Analyses of Redox-Active Iron Minerals: A Review of Nonmediated and Mediated Approaches. *Environ. Sci. Technol.* **2015**, *49* (10), 5862–5878. https://doi.org/10.1021/acs.est.5b00006.

(42) Wang, Z.; Yang, D.; Sham, T.-K. Effect of Oxidation State of Manganese in Manganese Oxide Thin Films on Their Capacitance Performances. *Surf. Sci.* **2018**, *676*, 71–76. https://doi.org/10.1016/j.susc.2017.12.011.

(43) Su, H.-Y.; Gorlin, Y.; C. Man, I.; Calle-Vallejo, F.; K. Nørskov, J.; F. Jaramillo, T.; Rossmeisl, J. Identifying Active Surface Phases for Metal Oxide Electrocatalysts: A Study of Manganese Oxide Bi-Functional Catalysts for Oxygen Reduction and Water Oxidation Catalysis. *Phys. Chem. Chem. Phys.* **2012**, *14* (40), 14010–14022. https://doi.org/10.1039/C2CP40841D.

(44) Min, X.; Chen, Y.; Kanan, M. W. Alkaline O2 Reduction on Oxide-Derived Au: High Activity and 4e− Selectivity without (100) Facets. *Phys. Chem. Chem. Phys.* **2014**, *16* (27), 13601–13604. https://doi.org/10.1039/C4CP01337A.

(45) Buvat, G.; Eslamibidgoli, M. J.; Youssef, A. H.; Garbarino, S.; Ruediger, A.; Eikerling, M.; Guay, D. Effect of IrO$_6$ Octahedron Distortion on the OER Activity at (100) IrO$_2$ Thin Film. *ACS Catal.* **2020**, *10* (1), 806–817. https://doi.org/10.1021/acscatal.9b04347.

(46) Zheng, Y.; Gao, R.; Zheng, L.; Sun, L.; Hu, Z.; Liu, X. Ultrathin Co$_3$O$_4$ Nanosheets with Edge-Enriched {111} Planes as Efficient Catalysts for Lithium–Oxygen Batteries. *ACS Catal.* **2019**, *9* (5), 3773–3782. https://doi.org/10.1021/acscatal.8b05182.

(47) Poulain, R.; Klein, A.; Proost, J. Electrocatalytic Properties of (100)-, (110)-, and (111)-Oriented NiO Thin Films toward the Oxygen Evolution Reaction. *J. Phys. Chem. C* **2018**, *122* (39), 22252–22263. https://doi.org/10.1021/acs.jpcc.8b05790.

(48) Maiyalagan, T.; Chemelewski, K. R.; Manthiram, A. Role of the Morphology and Surface Planes on the Catalytic Activity of Spinel $LiMn_{1.5}Ni_{0.5}O_4$ for Oxygen Evolution Reaction. *ACS Catal.* **2014**, *4* (2), 421–425. https://doi.org/10.1021/cs400981d.

(49) Mishra, R. K.; Thomas, G. Surface Energy of Spinel. *J. Appl. Phys.* **1977**, *48* (11), 4576–4580. https://doi.org/10.1063/1.323486.

(50) Moyer, J. A.; Gao, R.; Schiffer, P.; Martin, L. W. Epitaxial Growth of Highly-Crystalline Spinel Ferrite Thin Films on Perovskite Substrates for All-Oxide Devices. *Sci. Rep.* **2015**, *5* (1), 10363. https://doi.org/10.1038/srep10363.

(51) Zheng, H.; Straub, F.; Zhan, Q.; Yang, P.-L.; Hsieh, W.-K.; Zavaliche, F.; Chu, Y.-H.; Dahmen, U.; Ramesh, R. Self-Assembled Growth of $BiFeO_3$–$CoFe_2O_4$ Nanostructures. *Adv. Mater.* **2006**, *18* (20), 2747–2752. https://doi.org/10.1002/adma.200601215.

(52) West, A. R. *Solid State Chemistry and Its Applications*; John Wiley & Sons, 2014.

(53) Zhao, S.; Guo, T.; Fan, J.; Wang, L.; Han, M.; Wang, J.; Wu, Y.; Chen, Y. Versatile Synthesis of Ultrafine Ternary Spinel Oxides/Carbon Nanohybrids toward the Oxygen Reduction Reaction. *Energy Fuels* **2020**, *34* (7), 9069–9075. https://doi.org/10.1021/acs.energyfuels.0c01901.

(54) Wu, X.; Niu, Y.; Feng, B.; Yu, Y.; Huang, X.; Zhong, C.; Hu, W.; Li, C. M. Mesoporous Hollow Nitrogen-Doped Carbon Nanospheres with Embedded $MnFe_2O_4$/Fe Hybrid Nanoparticles as Efficient Bifunctional Oxygen Electrocatalysts in Alkaline Media. *ACS Appl. Mater. Interfaces* **2018**, *10* (24), 20440–20447. https://doi.org/10.1021/acsami.8b04012.

(55) Celorrio, V.; Dann, E.; Calvillo, L.; Morgan, D. J.; Hall, S. R.; Fermin, D. J. Oxygen Reduction at Carbon-Supported Lanthanides: The Role of the B-Site. *ChemElectroChem* **2016**, *3* (2), 283–291. https://doi.org/10.1002/celc.201500440.

(56) Liang, Y.; Li, Y.; Wang, H.; Zhou, J.; Wang, J.; Regier, T.; Dai, H. $Co_3O_4$ Nanocrystals on Graphene as a Synergistic Catalyst for Oxygen Reduction Reaction. *Nat. Mater.* **2011**, *10* (10), 780–786. https://doi.org/10.1038/nmat3087.

(57) Wang, L.; Sofer, Z.; Pumera, M. Will Any Crap We Put into Graphene Increase Its Electrocatalytic Effect? *ACS Nano* **2020**, *14* (1), 21–25. https://doi.org/10.1021/acsnano.9b00184.

(58) Kocha, S. S.; Zack, J. W.; Alia, S. M.; Neyerlin, K. C.; Pivovar, B. S. Influence of Ink Composition on the Electrochemical Properties of Pt/C Electrocatalysts. *ECS Trans.* **2013**, *50* (2), 1475. https://doi.org/10.1149/05002.1475ecst.

(59) Chen, Y.; Zhong, Q.; Li, G.; Tian, T.; Tan, J.; Pan, M. Electrochemical Study of Temperature and Nafion Effects on Interface Property for Oxygen Reduction Reaction. *Ionics* **2018**, *24* (12), 3905–3914. https://doi.org/10.1007/s11581-018-2533-3.

Table of Contents Figure:

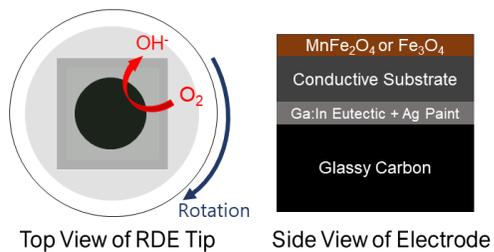

**Oxygen Reduction Electrocatalysis with Epitaxially Grown Spinel MnFe$_2$O$_4$ and Fe$_3$O$_4$**

Alexandria R. C. Bredar[†a], Miles D. Blanchet[‡a], Andricus R. Burton[†], Bethany Matthews[§], Steven R. Spurgeon[§], Ryan B. Comes[‡*], Byron H. Farnum[†*]

[†]Department of Chemistry and Biochemistry, Auburn University, Auburn, AL 36849
[‡]Department of Physics, Auburn University, Auburn, AL 36849
[§]Energy and Environment Directorate, Pacific Northwest National Laboratory, Richland, WA 99352

**Supplementary Information**

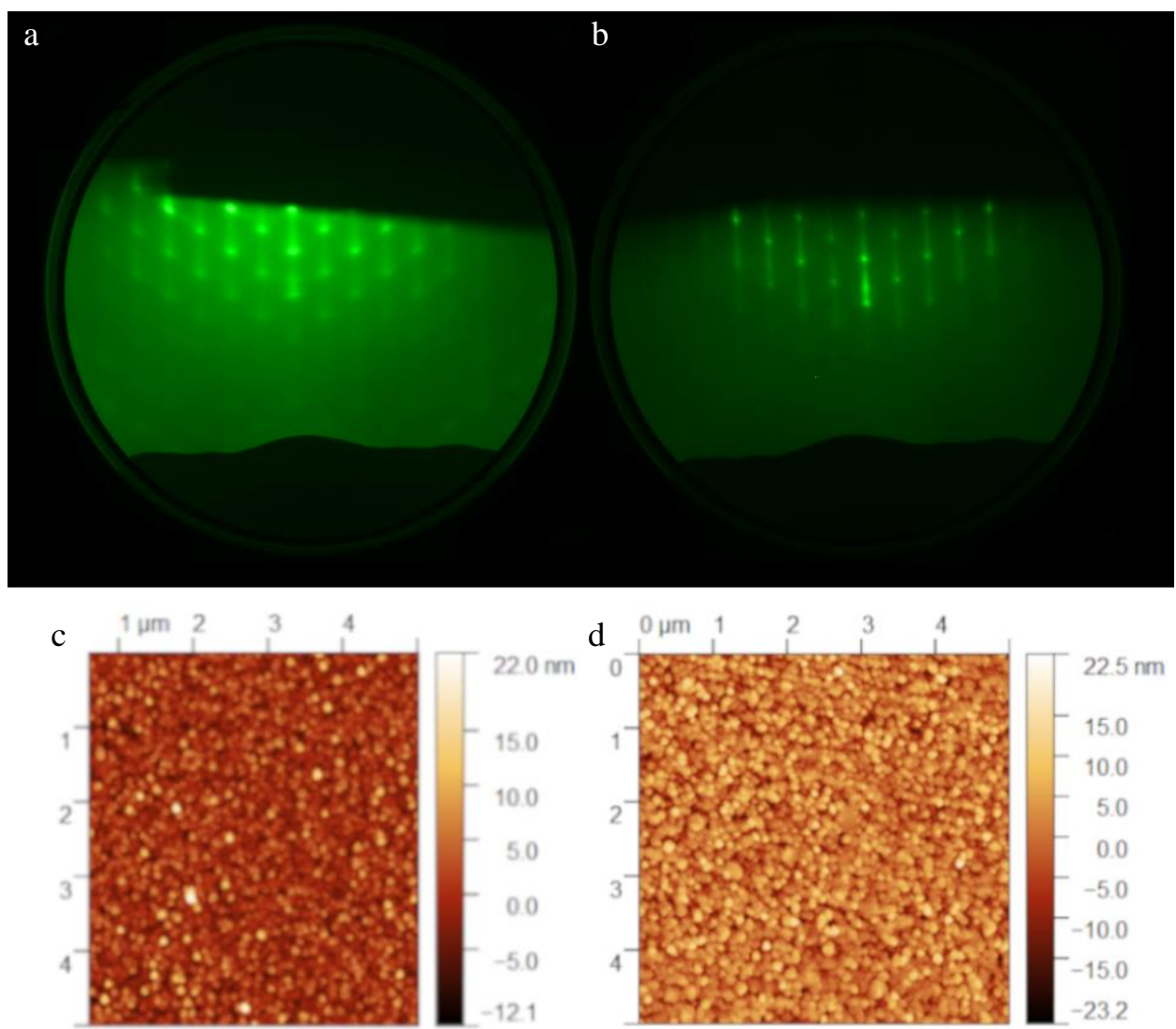

**Figure S1.** RHEED and AFM images of (a,c) 16 nm (001) $MnFe_2O_4$ and (b,d) 21 nm (001) $Fe_3O_4$, with RMS roughness of this $MnFe_2O_4$ and $Fe_3O_4$ determined as $3.6 \pm 0.5$ nm and $5.3 \pm 0.5$ nm, respectively.

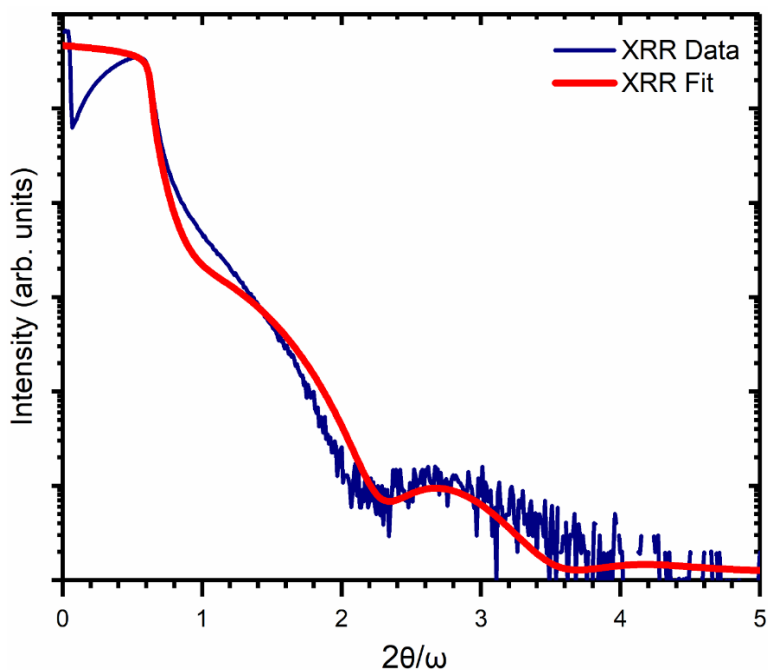

**Figure S2.** XRR fit of 6 nm (111) MnFe$_2$O$_4$, used in determining sample thicknesses. Uncertainty is within 1 nm.

**Table S1: MnFe$_2$O$_4$ atomic% from EDS**

| EDS | 6 nm (111) MnFe$_2$O$_4$ | | 6 nm (001) MnFe$_2$O$_4$ | | 16 nm (001) MnFe$_2$O$_4$ | |
|---|---|---|---|---|---|---|
| **Film Region** | **Mn** | **Fe** | **Mn** | **Fe** | **Mn** | **Fe** |
| 1 | 29.4 | 70.6 | 8.0 | 92.0 | 29.9 | 70.1 |
| 2 | 40.1 | 59.9 | 34.9 | 65.1 | 28.6 | 72.3 |
| 3 | 31.8 | 68.2 | 22.4 | 77.6 | 27.7 | 71.4 |
| 4 | 33.9 | 66.1 | 38.6 | 61.4 | 29.4 | 70.6 |
| Avg | 33.8 ± 4.6 | 66.2 ± 4.6 | 26.0 ± 13.8 | 74.0 ± 13.8 | 28.9 ± 1.0 | 71.1 ± 1.0 |
| **Mn/Fe** | **0.51 ± 0.15** | | **0.31 ± 0.56** | | **0.41 ± 0.04** | |

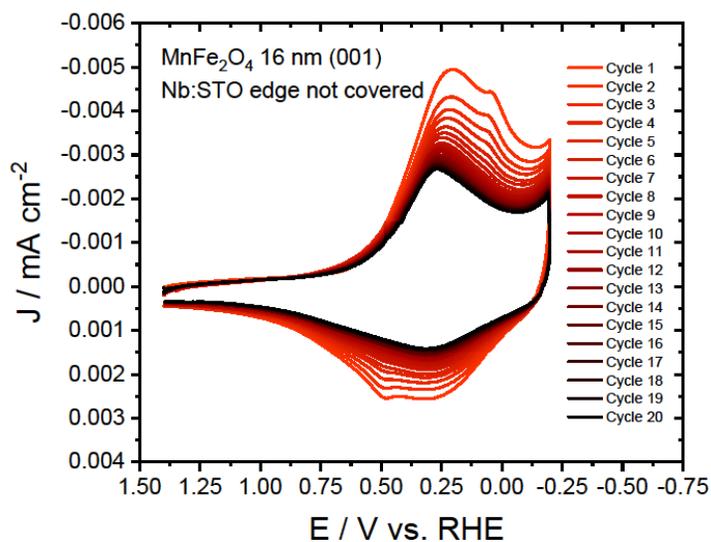

**Figure S3.** Electrochemical cycling of 16 nm (001) MnFe$_2$O$_4$ with underlying Nb:STO substrate exposed, showing film degradation.

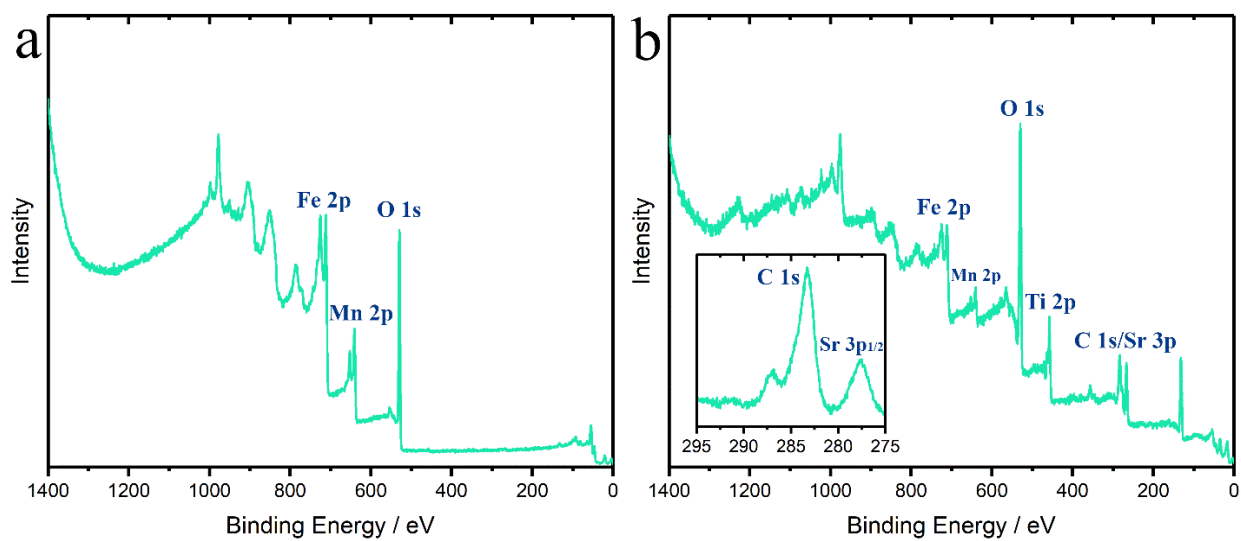

**Figure S4.** XPS Surveys of 16 nm (001) MnFe$_2$O$_4$ (a) immediately after growth, under vacuum and (b) after catalysis experiments. Post-experiment XPS reveals Ti and Sr from the Nb:STO substrate indicating etching of the film.

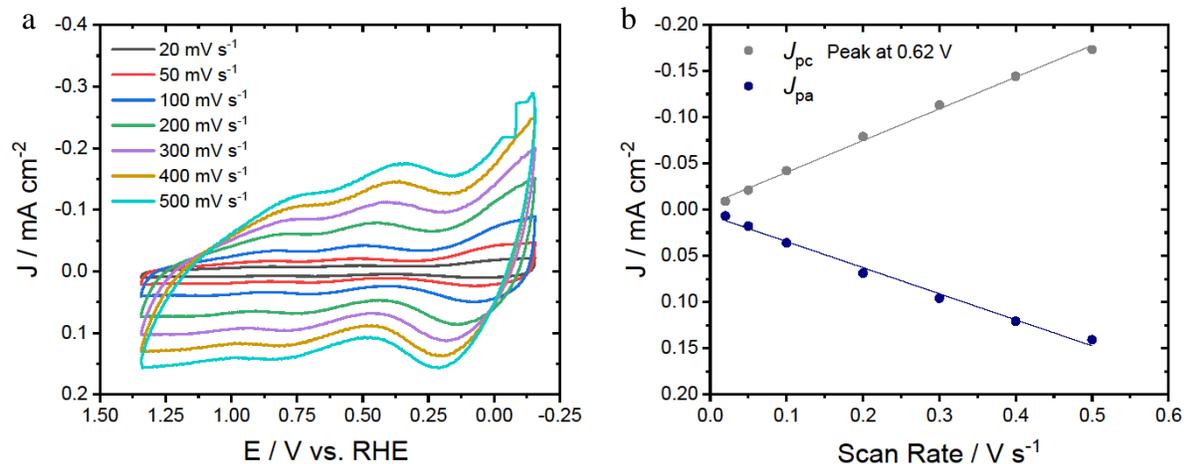

**Figure S5.** a) Scan rate normalized CVs of MnFe$_2$O$_4$ at 20, 50, 100, 200, 300, 400, and 500 mV s$^{-1}$ b) $i_{pc}$ and $i_{pa}$ of the redox wave at 0.62 V vs RHE demonstrating a linear dependence and scan rate and thus a surface bound redox feature.

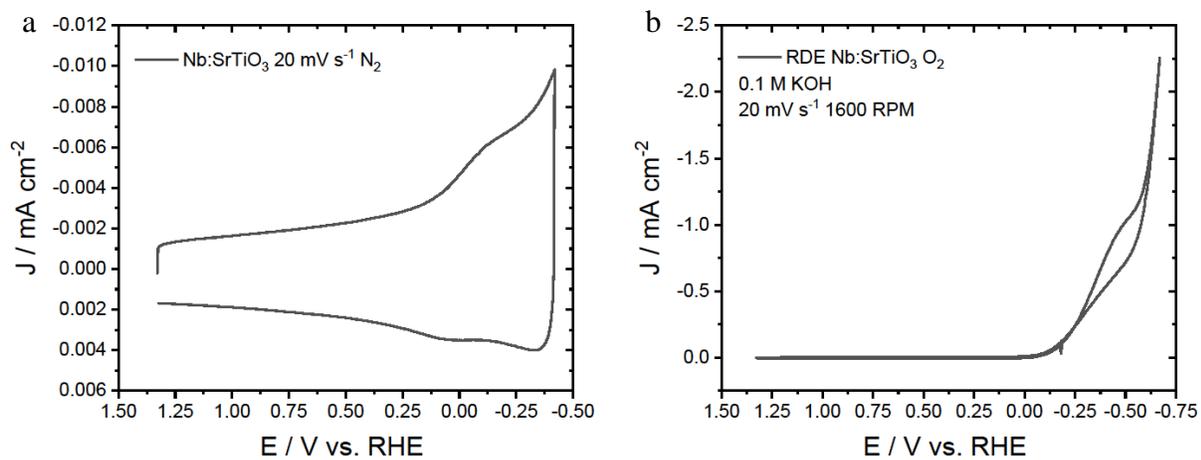

**Figure S6.** CV of Nb:STO substrate under a) N$_2$ and b) RDE of Nb:STO substrate under O$_2$ at 1600 rpm

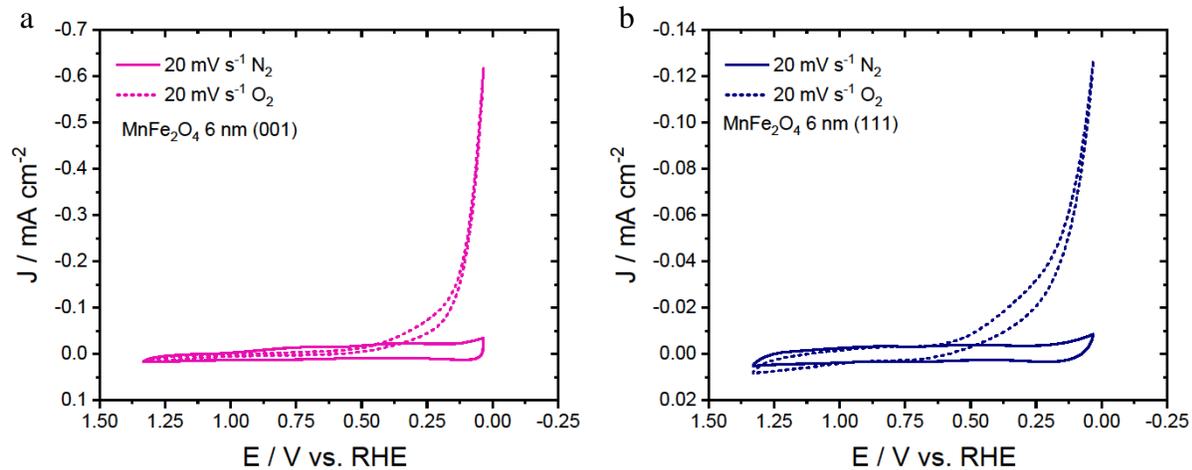

**Figure S7.** CV of 6 nm (001) and (111) films before rotation experiments comparing $N_2$ and $O_2$

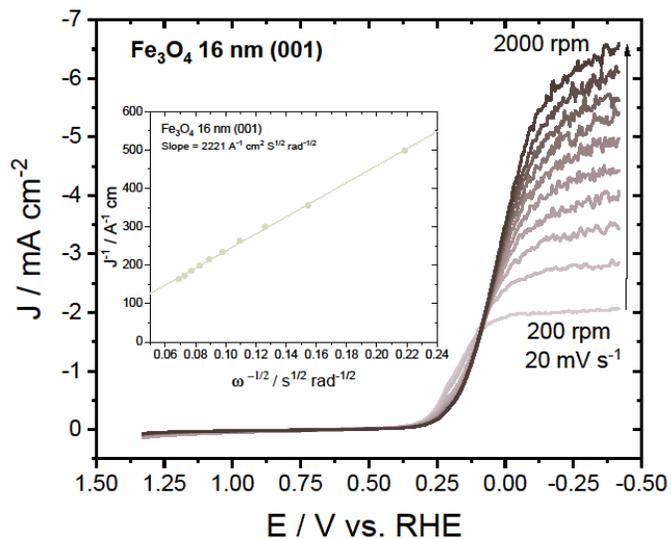

**Figure S8.** RDE data of $Fe_3O_4$ 21 nm (001) film

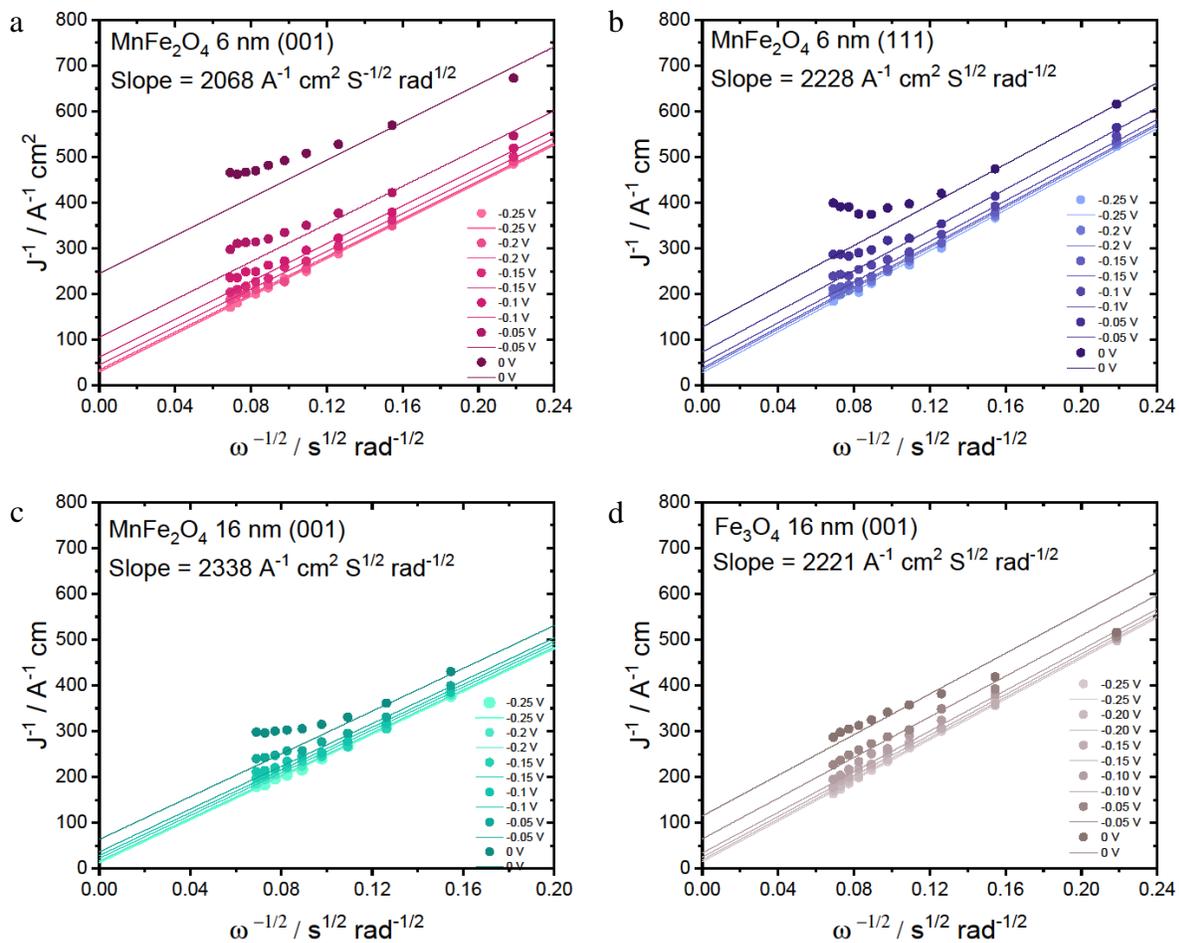

**Figure S9.** K-L plots for a) MnFe$_2$O$_4$ 6 nm (001) b) MnFe$_2$O$_4$ 6 nm (111) c) MnFe$_2$O$_4$ 16 nm (001) d) Fe$_3$O$_4$ 21 nm (001).

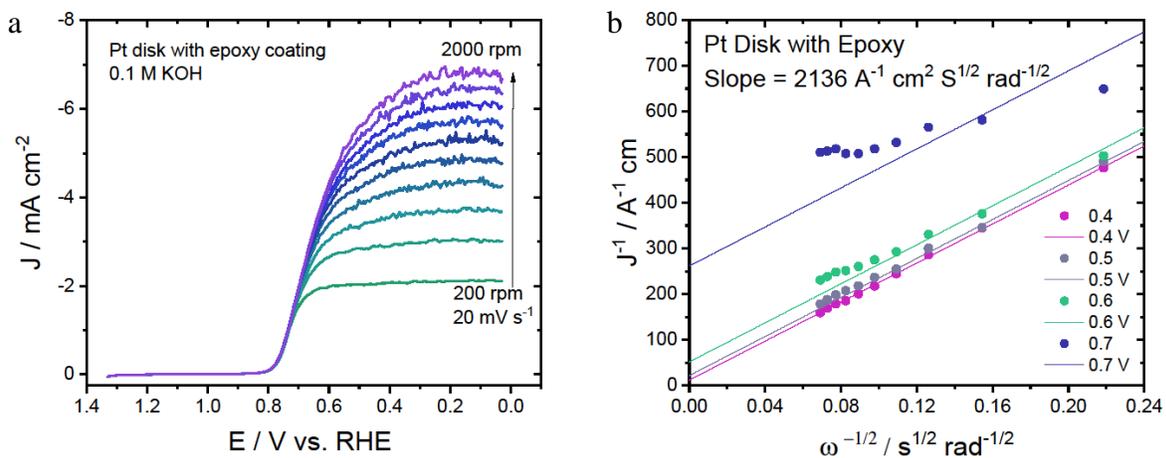

**Figure S10.** a) RDE and b) K-L plots for Pt disk control with epoxy ring as a control for MBE electrodes.

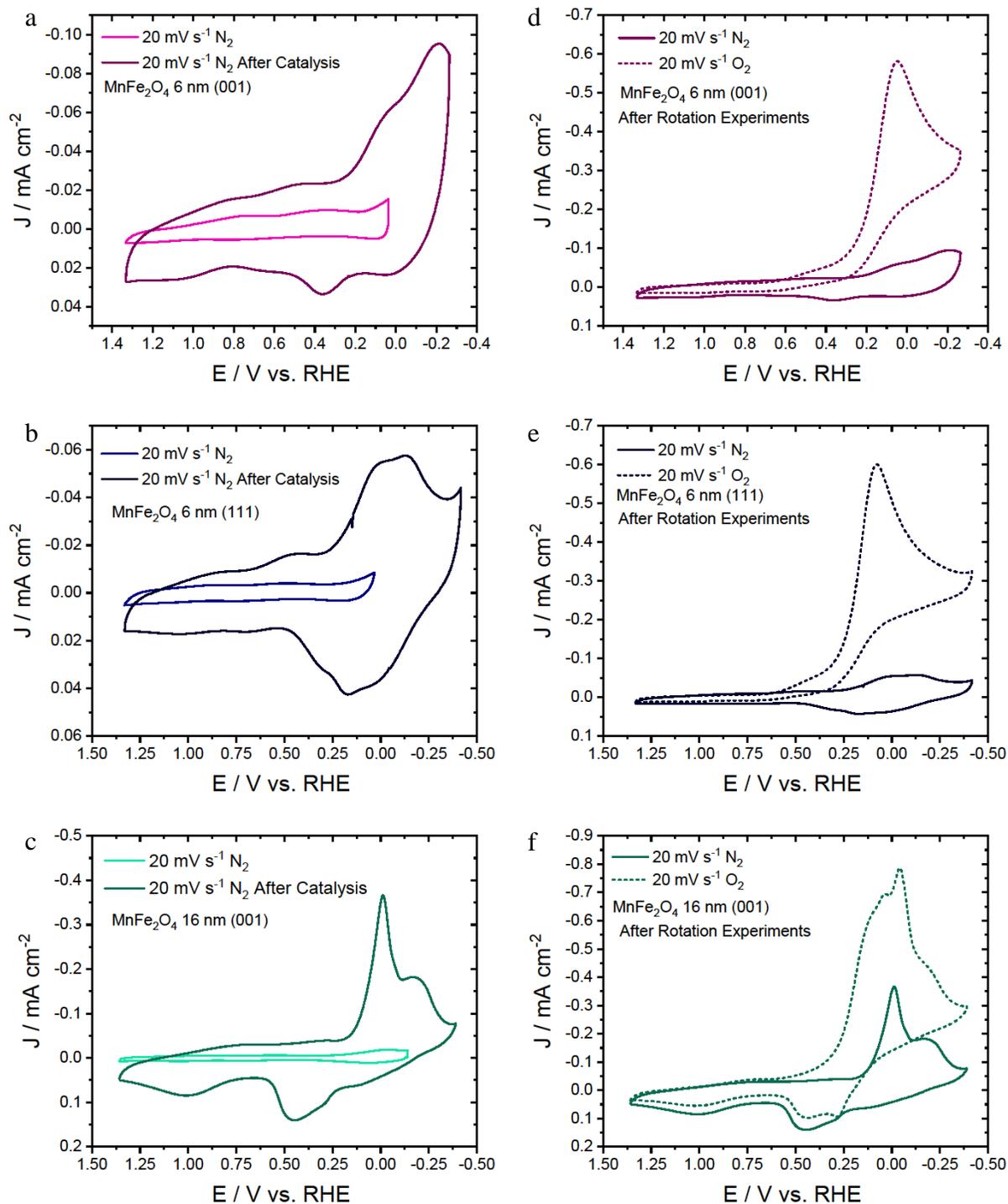

**Figure S11.** $N_2$ before and after catalysis CV comparison for a) 6 nm (001), b) 6 nm (111), and c) 16 nm (001) films. $N_2$ and $O_2$ after catalysis CV comparison for d) 6 nm (001), e) 6 nm (111), and f) 16 nm (001) films.

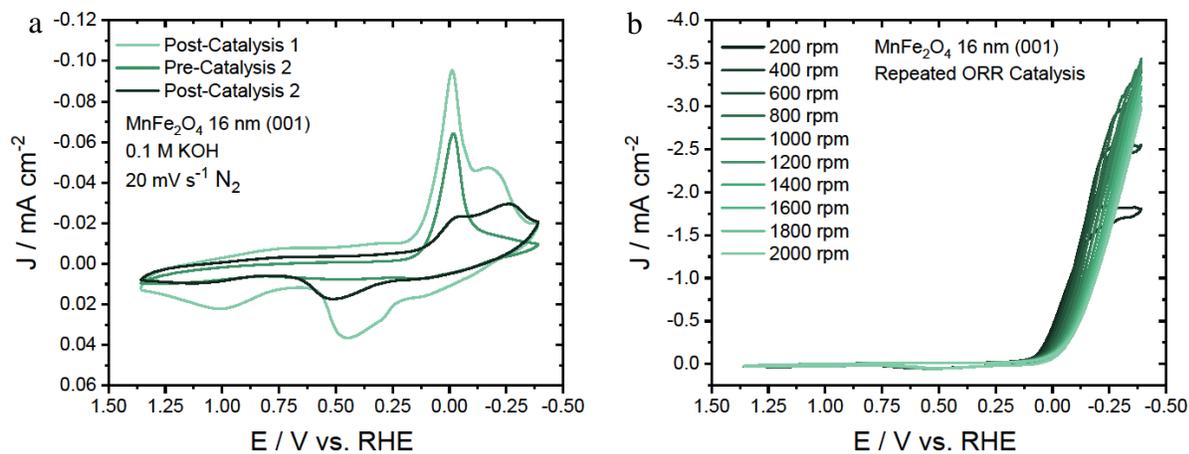

**Figure S12.** a) CV and b) RDE of MnFe$_2$O$_4$ 16 nm (001) of repeated ORR electrocatalysis on the same film

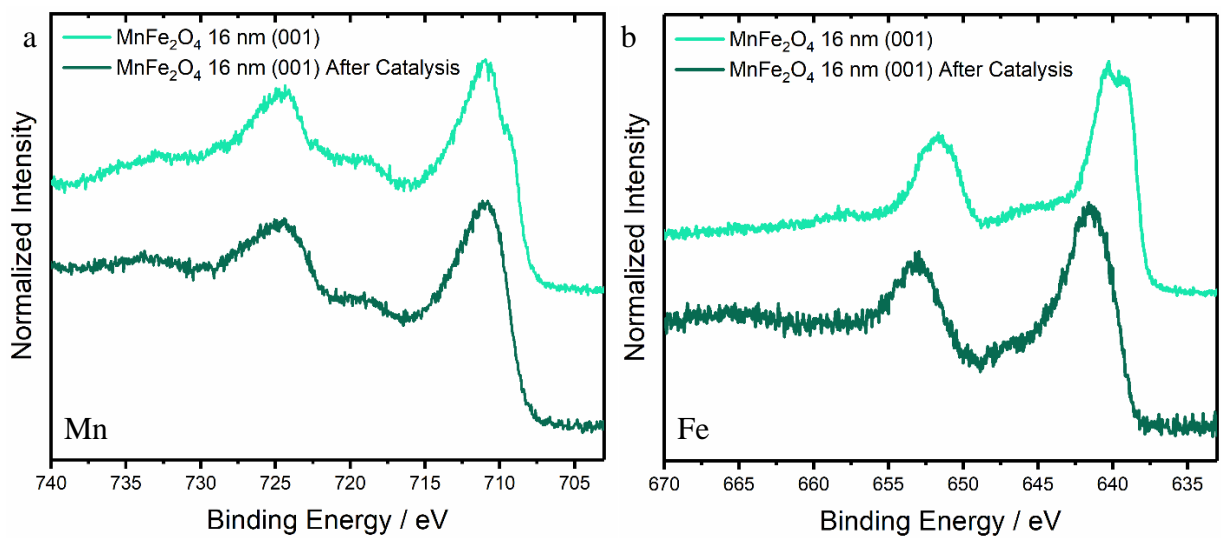

**Figure S13.** Before atmosphere exposure and after-catalysis XPS spectra of MnFe$_2$O$_4$ 16 nm (001) for (a) Mn region and (b) Fe region.

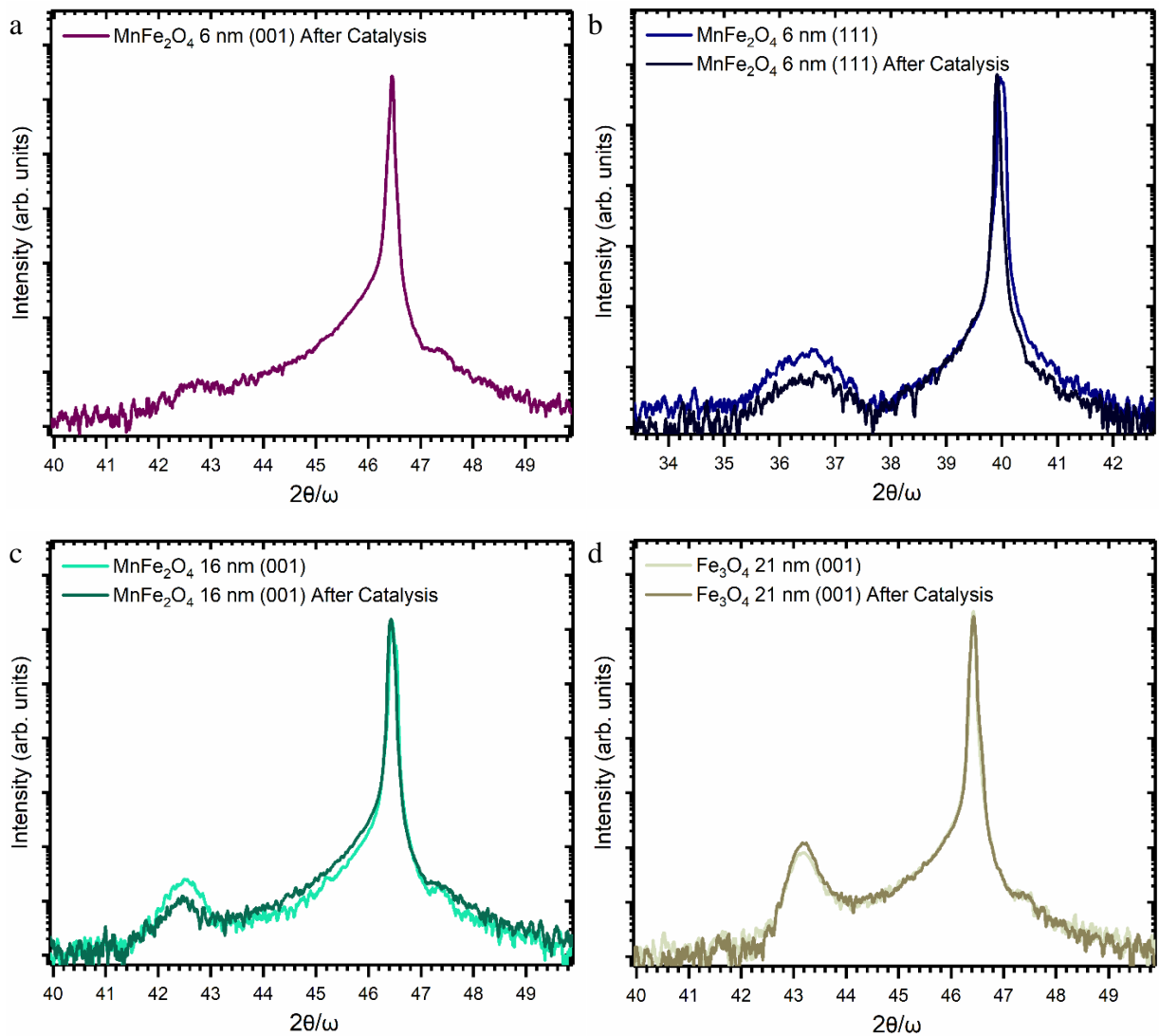

**Figure S14.** HRXRD diffractograms of samples after catalysis experiments were performed of (a) MnFe$_2$O$_4$ 6 nm (001), (b) MnFe$_2$O$_4$ 6 nm (111), (c) MnFe$_2$O$_4$ 16 nm (001) and (d) Fe$_3$O$_4$ 21 nm (001). Due to the small volume and rough surface quality of MnFe$_2$O$_4$ 6 nm (001), the film peak is convolved with the shoulder of Nb:STO's (002) peak.

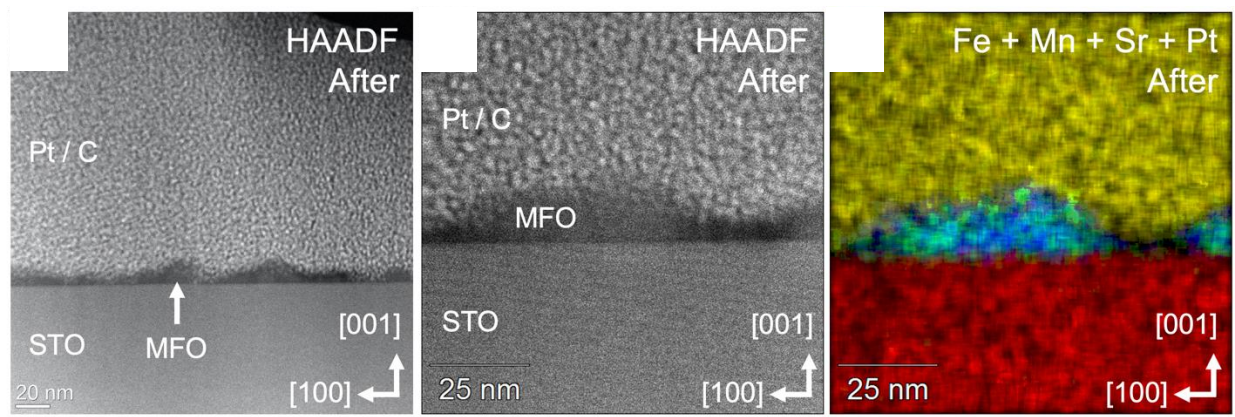

**Figure S15.** STEM analysis of MnFe$_2$O$_4$ 6 nm (001) after catalysis.

## Calculation of Total Charge Density of {111} Surface

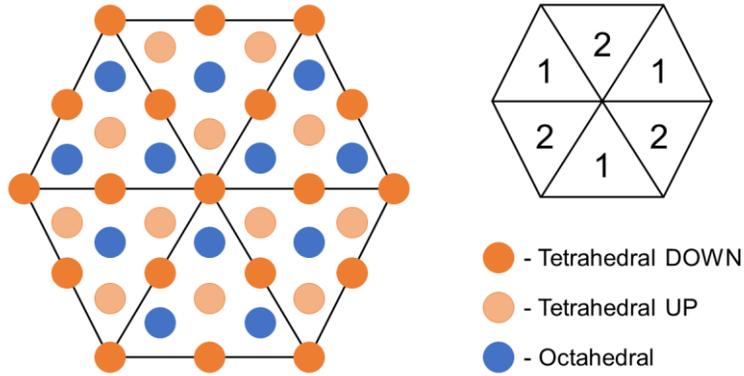

**Figure S16.** Projected {111} surface showing adjacent (111) planes from six unit cells to emphasize the atomic sharing between unit cells. Each edge of the overall hexagon is equal to 12.02 Å = $a\sqrt{2}$, where $a$ = 8.50 Å is the lattice parameter. The {111} plane is divided into two types of triangular faces which differ in their tetrahedral and octahedral occupancy. These are labeled (1) and (2) and alternate throughout the plane.

Area of {111} face per unit cell
Edge length, c = 12.02 Å
Area of equilateral triangle = $\frac{c^2\sqrt{3}}{4}$ = 62.6 Å² = 62.6 x $10^{-16}$ cm²

Number of atoms on {111} face per unit cell
Two types of {111} triangular faces – (1) and (2)

Face (1)
Tetrahedral Sites: 3 corner sites (1/6 shared), 3 edge sites (1/2 shared), 1 body = 3 total
Octahedral Sites: 3 body = 3 total
6 total atoms = 3 tetrahedral sites + 3 octahedral sites

Face (2)
Tetrahedral Sites: 3 corner sites (1/6 shared), 3 edge sites (1/2 shared), 3 body = 5 total
Octahedral Sites: 1 body = 1 total
6 total atoms = 5 tetrahedral sites + 1 octahedral site

Face(1) + Face (2) = 8 tetrahedral sites and 4 octahedral sites = 12 total over two unit cells

Normal spinel – Mn in all tetrahedral sites = 8/12 = 67% Mn site density
Inverse spinel – Mn in all octahedral sites = 4/12 = 33% Mn site density
20% inversion – Mn in 80% tetrahedral sites and 20% octahedral sites = 0.8*(8/12) + 0.2*(4/12) = 60% Mn site density

Charge density of {111} face
Given that Face (1) and Face (2) both contain 6 atoms, the charge density will be the same for both faces, regardless of coordination sites, if we assume that both sites can be oxidized/reduced by 1e$^-$ (*i.e.* Fe$^{III}$/Fe$^{II}$ or Mn$^{III}$/Mn$^{II}$)

$$\left(\frac{6 \ atoms}{1}\right)\left(\frac{mol}{6.022 \times 10^{23} \ atoms}\right) = 9.963 \times 10^{-24} \ mol$$

$$\left(\frac{9.963 \times 10^{-24} \ mol}{1}\right)\left(\frac{96,485 \ C}{mol}\right)\left(\frac{1}{62.6 \times 10^{-16} \ cm^2}\right) = 154 \ \mu C/cm^2$$

Charge density calculations
The areas of anodic peaks and cathodic peaks were measured with Pine Aftermath software and reported in **Table S2** in units of nW. These areas represent the total area under both peaks (*i.e.* E$_{1/2}$ = 0.92 and 0.65 V). Anodic and cathodic areas were then average to give a final area which was converted to charge density by the equation below. A$_{geo}$ is the geometric area of the exposed MnFe$_2$O$_4$ surface and *v* is the scan rate of the cyclic voltammetry experiment. Percent Mn coverage was calculated by dividing the measured charge density by the total theoretical charge density derived above.

$$charge \ density \ (\mu C/cm^2) = \left(\frac{X \ nW}{1}\right)\left(\frac{W}{10^9 \ nW}\right)\left(\frac{C \ V}{W \ s}\right)\left(\frac{s}{v \ V}\right)\left(\frac{1}{A_{geo} \ cm^2}\right)\left(\frac{10^6 \ \mu C}{C}\right)$$

**Table S2.** Summary of integrated peak areas and charge density calculations for Mn-based redox features under N$_2$

|  | 6 nm (001) | 6 nm (111) | 16 nm (001) |
|---|---|---|---|
| A$_{geo}$ / cm$^2$ | 0.038 | 0.034 | 0.030 |
| *v* / V s$^{-1}$ | 0.02 | 0.02 | 0.02 |
| Cathodic Area / nW | 15.0 | 6.5 | 18.8 |
| Anodic Area / nW | 14.5 | 4.3 | 18.1 |
| Average Area / nW | 14.7 | 5.4 | 18.5 |
| Charge Density / μC cm$^{-2}$ | 21.7 | 9.0 | 24.3 |
| Percent Mn Coverage | 14.1% | 5.8% | 15.8% |